\documentclass[twocolumn,american,aps,amsmath,amssymb,floatfix]{revtex4}
\usepackage[latin9]{inputenc}
\usepackage{graphicx}
\usepackage{esint}

\makeatletter
\@ifundefined{textcolor}{}
{%
 \definecolor{BLACK}{gray}{0}
 \definecolor{WHITE}{gray}{1}
 \definecolor{RED}{rgb}{1,0,0}
 \definecolor{GREEN}{rgb}{0,1,0}
 \definecolor{BLUE}{rgb}{0,0,1}
 \definecolor{CYAN}{cmyk}{1,0,0,0}
 \definecolor{MAGENTA}{cmyk}{0,1,0,0}
 \definecolor{YELLOW}{cmyk}{0,0,1,0}
}

\usepackage{bm}
\newtheorem{condition}{Condition}
\newtheorem{definition}{Definition}
\newtheorem{theorem}{Theorem}

\makeatother

\usepackage{babel}
\begin{document}

\title{A general concept of natural information equilibrium:\\
from the ideal gas law to the K-Trumpler effect}

\author{P. Fielitz}

\email{peter.fielitz@tu-clausthal.de}

\selectlanguage{american}%

\author{G. Borchardt}

\email{guenter.borchardt@tu-clausthal.de}

\selectlanguage{american}%

\affiliation{Technische Universität Clausthal, Fakultät für Natur- und Materialwissenschaften,\\
 Robert-Koch-Str.~42, D-38678 Clausthal-Zellerfeld, Germany }

\date{\today}
\begin{abstract}
Information theory provides shortcuts which allow one to deal with
complex systems. The basic idea one uses for this purpose is the maximum
entropy principle developed by Jaynes. However, an extension of this
maximum entropy principle to systems far from thermodynamic equilibrium
or even to non-physical systems is problematic because it requires
an adequate choice of constraints. In this paper we discuss a general
concept of natural information equilibrium which does not require
any choice of adequate constraints. It is, therefore, directly applicable
to systems far from thermodynamic equilibrium and to non-physical
systems/processes (e.g. biological processes and economical processes).
We demonstrate the validity and the applicability of the concept by
three well understood physical processes. As an interesting astronomical
application we will show that the concept of natural information equilibrium
allows one to rationalize and to quantify the K-Trumpler effect. 
\end{abstract}
\maketitle

\section{Introduction}

In 2009 we published two versions of an information transfer model
of natural processes in arXiv.org (0905.0610v1,v2). These preliminary
versions presented first ideas how one can model information transfer
in natural systems. A more mature version of such an information transfer
concept was published in 2011 in Physics Essays \cite{Fielitz2011}.
In 2013 the arXiv paper has been completely rewritten (0905.0610v3)
and became based on reference \cite{Fielitz2011}. At about the same
time as we prepared the update, Smith \cite{Smith2013} started an
interesting blog (what he calls a working paper) in which he applies
the proposed information transfer model to economical processes. To
our knowledge it is the first application of the information transfer
model to non-physical systems/processes.

In the meantime the main result of this arXiv paper (the phenomenological
description of the K-Trumpler effect) is accepted for publication
in Physics Essays \cite{Fielitz2014}. Readers who are mainly interested
in the phenomenological description of the K-Trumpler effect should
also read reference \cite{Fielitz2014}. The K-Trumpler effect is,
however, only one (non-trivial) application of the proposed generalized
concept of information transfer \cite{Fielitz2011}. The most interesting
feature of the information transfer concept is its \emph{system/process
independent character} so that it can be helpful to get a new (phenomenological)
point of view on a controversial phenomenon (as demonstrated for the
Titius-Bode rule \cite{Fielitz2011} and the K-Trumpler effect \cite{Fielitz2014}).
This means that this arXiv paper remains interesting for readers who
want to apply the proposed information transfer concept to their own
problems.

In this fourth version (0905.0610v4) we have expanded the introduction
to provide some information about our motivation to develop a generalized
concept of information transfer \cite{Fielitz2011}. Since version
v3 of this arXiv paper we consider only the case of information equilibrium
so that we changed also the title of this version (v4) to make explicitly
clear that we focus here on a general concept of natural information
equilibrium. The universality of this concept becomes especially clear
if one compares equations (\ref{eqv4_40}) and (\ref{eqv4_41}) where
the physical concept of force equilibrium is directly compared with
the general concept of information equilibrium. The most important
definition of such a concept is an appropriate definition of a \emph{natural}
amount of information. We explain in this version why the defined
amount of information \cite{Fielitz2011}, which is related to a generic
process variable (see equation (\ref{eqv3_01})), is a natural amount
of information (see appendix \ref{sec:Appendix-Signals}). This is
an important update in the notation because in references \cite{Fielitz2011}
and \cite{Fielitz2014} the defined amount of information was not
yet explicitly called ``natural''. In chapter \ref{sec:Phys-interpretation}
we demonstrate by two physical examples that the natural amount of
information is not an abstract number but has real physical meaning
-- if the internal details of a process are well understood. Further
improvements in this version are:\\
(i) The age of a star is now more precisely defined (see beginning
of chapter \ref{sec:Photon-propagation}).\\
(ii) We present a more direct application of the information transfer
concept to rationalize the K-Trumpler effect and thus avoid the introduction
of an abstract detector (see section \ref{sub:Direct-application}).
This approach is new and has not been presented in reference \cite{Fielitz2014}.

\subsection{Motivation to develop a generalized information transfer concept}

In materials science one important topic is self-diffusion of the
constituent elements in solid materials. Materials scientists, like
the authors (physics and chemistry of materials), are very confident
of their understanding of self-diffusion in homogeneous solids. However,
the following thought experiment confronted us with the fact that
we did \emph{not} (yet) understand \emph{all aspects} of the diffusion
process.

\subsubsection{A thought experiment which was totally incomprehensible for materials
scientists\label{sub:Thought-experiment}}

\begin{figure}
\includegraphics[width=7.5cm]{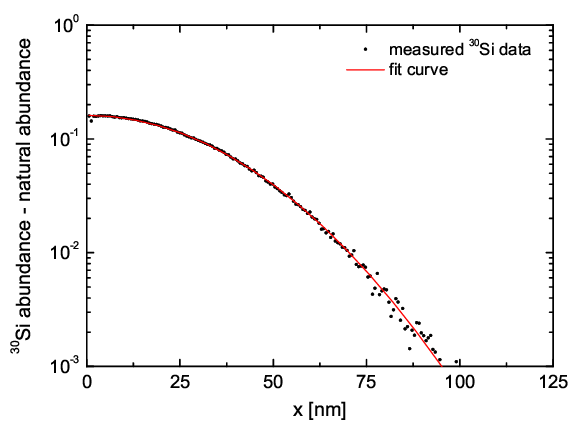} \protect\caption{$^{30}{\rm Si}$ depth distribution in single crystalline 2/1-mullite
measured by SIMS depth profiling after the diffusion annealing at
a given temperature \cite{Fielitz2003}. The natural abundance of
the $^{30}{\rm Si}$ isotope has been subtracted from the data.}

\label{fig:00} 
\end{figure}

The authors measure self-diffusion in solid materials and apply for
this task natural tracer isotopes which are deposited as thin tracer
layers on samples in which the diffusivity of a tracer isotope has
to be measured. For example in the work of reference \cite{Fielitz2003}
a thin aluminosilicate layer enriched in $^{30}{\rm Si}$ (a natural
silicon isotope) was sputter deposited on the surface of single crystalline
2/1-mullite (${\rm 2Al}_{{\rm 2}}{\rm O}_{{\rm 3}}\cdot{\rm SiO}_{{\rm 2}}$)
samples. After the diffusion annealing of the samples the depth distribution
of the natural Si isotopes can be measured by SIMS depth profiling.
We will not go into the details of the measurement technique. For
the reader it is only important to see that one obtains experimental
tracer depth profiles as shown in Fig.~\ref{fig:00}. 

The black points in Fig.~\ref{fig:00} are measured data of the $^{30}{\rm Si}$
isotope distribution after the diffusion annealing at a given temperature.
The natural abundance of the $^{30}{\rm Si}$ isotope has been subtracted
from the measured data. To evaluate the silicon tracer diffusion coefficient
from this data the thin layer solution of the diffusion equation for
a semi-infinite medium can be applied (\cite{Crank1975} p. 13) 
\begin{equation}
C(x,t)=\frac{M}{\left(\pi D\, t\right)^{1/2}}\exp\left(-\frac{x^{2}}{4D\, t}\right)\label{eqv4_01}
\end{equation}
where $D=D(T)$ is the temperature dependent tracer diffusion coefficient,
$t$ the annealing time, and $M$ the amount of diffusing tracer isotopes
deposited at $t=0$ in the plane $x=0$. The red solid line in Fig.~\ref{fig:00}
shows a least squares fit of equation (\ref{eqv4_01}). It is this
good agreement between diffusion theory (equation (\ref{eqv4_01}))
and measured data (Fig.~\ref{fig:00}) which makes materials scientists
confident that one understands \emph{all aspects} of the self-diffusion
in homogeneous solids -- until one makes a thought experiment as discussed
now. Let us assume in our thought experiment that we could build a
device which locally measures the concentration, $C(x,t)$, of the
tracer isotopes in a thin plane and its flux, $j_{x}(x,t)$, through
this plane. As output our considered device then delivers the local
ratio between flux and concentration 
\begin{equation}
\left(\frac{j_{x}}{C}\right)_{{\rm device}}\label{eqv4_02}
\end{equation}
It is a virtual measurement device in a thought experiment because
there is no technical realization (at least to date) of such a measurement
device in a solid medium. The local tracer particle flux is defined
by 
\begin{equation}
j_{x}(x,t)=\frac{1}{A_{0}}\frac{\partial N(x,t)}{\partial t}\label{eqv4_03}
\end{equation}
where $\partial N$ is the infinitesimal number of tracer particles
which pass through the unit area $A_{0}$ in $x$ direction during
the infinitesimal time interval $\partial t$. The local concentration
of the tracer particles is given by definition as 
\begin{equation}
C(x,t)=\frac{1}{A_{0}}\frac{\partial N(x,t)}{\partial x}\label{eqv4_04}
\end{equation}
If one considers the local ratio between flux and concentration one
gets from equations (\ref{eqv4_03}) and (\ref{eqv4_04}) the formal
relation 
\begin{equation}
\left(\frac{j_{x}}{C}\right)_{{\rm device}}=\frac{\partial x}{\partial t}\label{eqv4_05}
\end{equation}
so that the output value of the considered device will have the dimension
of a velocity. For materials scientists such a device would be a useful
device if it could simplify the measurement of the self diffusion
coefficient $D(T)$. That is, it was interesting to see how the output
value of such a virtual device depends on the diffusion coefficient.
From diffusion theory \cite{Crank1975} it is well known that the
tracer particle flux is given by (Fick's first law) 
\begin{equation}
j_{x}(x,t)=-D\frac{\partial C(x,t)}{\partial x}\label{eqv4_06}
\end{equation}
Combining equations (\ref{eqv4_01}), (\ref{eqv4_05}) and (\ref{eqv4_06})
one gets 
\begin{equation}
\left(\frac{j_{x}}{C}\right)_{{\rm device}}=\frac{\partial x}{\partial t}=\frac{1}{2}\frac{x}{t}\label{eqv4_07}
\end{equation}
As we derived equation (\ref{eqv4_07}) the very first time we were
astonished to see that equation (\ref{eqv4_07}) is totally independent
of the diffusion coefficient so that the considered virtual device
(\ref{eqv4_02}) has the strange property that its output value is
independent of the temperature and is independent of the properties
of the tracer particles and the diffusion medium. It is also very
strange that one could use such a device to measure (!) the diffusion
annealing time interval, $t$, at any stage after the start of the
diffusion experiment. With other words equation (\ref{eqv4_07}) was
completely incomprehensible for materials scientists and demonstrated
that there was \emph{some aspect} related to the self-diffusion process
in a homogeneous medium that we did not (yet) understand. This was
an unsatisfactory situation at that time as we conceived this thought
experiment.

The reader could argue that equation (\ref{eqv4_07}) is only correct
if one considers infinitesimal tracer layers. This argument is correct,
but materials scientists who have practical experience with tracer
diffusion experiments know that equation (\ref{eqv4_01}) is also
practically applicable for thick layers and even for constant diffusion
sources if one considers the diffusion process sufficiently far from
the tracer source, so that equation (\ref{eqv4_07}) becomes general
if one writes 
\begin{equation}
\left(\frac{j_{x}}{C}\right)_{{\rm device}}=\frac{\partial x}{\partial t}=\frac{1}{2}\frac{x}{t}\quad{\rm if}\quad x\ge x^{*}\label{eqv4_08}
\end{equation}
where $x^{*}$ is a distance which is sufficiently far from the tracer
source. The absolute value of $x^{*}$ depends on the deviation from
the ideal case (\ref{eqv4_07}) which can be practically accepted.
If the deviation is practically acceptable for $x=x^{*}$ the deviation
will be even smaller for $x\ge x^{*}$ if the diffusion time, $t$,
remains fixed (see \cite{Fielitz2011} where the case of a constant
diffusion source is discussed in detail). This is very well confirmed
practical experience from diffusion experiments.

If one does not understand a result (equation (\ref{eqv4_07}) or
(\ref{eqv4_08}), respectively) one tends to believe that it is accidental.
Furthermore, we could initially assume that this has been only a virtual
measurement device in a thought experiment which has no meaning at
all. Nevertheless, the situation remained unsatisfactory and we started
to look for a \emph{real} measurement device with a similar \emph{simple}
dependence of its output value from the process variables ($x$ and
$t$ in this case). In the next section we discuss a corresponding
real device.

\subsubsection{Looking from a new point of view on a mechanical gas pressure measurement
device\label{sub:Pressure-Meas-Device}}

Let us consider a mechanical gas pressure measurement device 
\begin{equation}
\left(\frac{\left|F\right|}{A}\right)_{{\rm device}}\label{eqv4_09}
\end{equation}
where $\left|F\right|$ is the measured force which acts on a membrane
with the area $A$. The force is induced by the gas particles (molecules
or atoms) which are impinging on the membrane. However, to measure
a force the membrane must be displaced by a small absolute value $\left|\partial l\right|$
so that we can also consider a small amount of energy $\left|\partial E\right|=\left|F\right|\left|\partial l\right|$.
This small amount of energy is transferred from the impinging gas
particles to the membrane and is converted to a small amount of mechanical
work. Considering the corresponding small volume element $\left|\partial V\right|=A\left|\partial l\right|$
one gets the following equation 
\begin{equation}
\left(\frac{\left|F\right|}{A}\right)_{{\rm device}}=\frac{\left|\partial E\right|}{\left|\partial V\right|}\label{eqv4_10}
\end{equation}
This equation corresponds to equation (\ref{eqv4_05}) and like in
the thought experiment above (section \ref{sub:Thought-experiment})
one can ask how the output value of the device depends on the process
variables ($E$ and $V$ in this case). From the thermodynamics of
the ideal gas one gets 
\begin{equation}
\left(\frac{\left|F\right|}{A}\right)_{{\rm device}}=\frac{\left|\partial E\right|}{\left|\partial V\right|}=\frac{2}{f}\frac{E}{V}\label{eqv4_11}
\end{equation}
where $E$ is the internal energy of the ideal gas, $V$ the ideal
gas volume and $f$ the degree of freedom per gas particle. Furthermore,
one knows from thermodynamics that a real gas becomes an ideal gas
if the gas box volume becomes sufficiently large so that the gas particles
can be considered as an ensemble of point particles with no form of
interaction. That is, equation (\ref{eqv4_11}) becomes general if
one writes 
\begin{equation}
\left(\frac{\left|F\right|}{A}\right)_{{\rm device}}=\frac{\left|\partial E\right|}{\left|\partial V\right|}=\frac{2}{f}\frac{E}{V}\quad{\rm if}\quad V\ge V^{*}\label{eqv4_12}
\end{equation}
where $V^{*}$ is a sufficiently large volume. It is well known experimental
experience that if a real gas can be practically approximated by the
ideal gas (equation (\ref{eqv4_11})) at $V=V^{*}$ it will even be
better approximated for volumes $V\ge V^{*}$ if the internal energy,
$E$, remains fixed (which implies constant temperature). If one compares
equations (\ref{eqv4_08}) and (\ref{eqv4_12}) it is quite surprising
that one gets mathematically equivalent equations for two different
systems with different process variables.

Seeing equations (\ref{eqv4_08}) and (\ref{eqv4_12}) it now became
harder to assume that its mathematical equivalence could be accidental
so that we started at that point to look seriously for a fundamental
law which could explain the occurrence of equations (\ref{eqv4_08})
and (\ref{eqv4_12}). It was clear that a law behind equations (\ref{eqv4_08})
and (\ref{eqv4_12}), and hence behind different systems, must be
system independent. Finally we found the required fundamental law
in information theory (see inequality \ref{eqv3_02}) and developed
a generalized information transfer concept of natural processes \cite{Fielitz2011}
as described briefly in chapter \ref{sec:Basics}. In chapter \ref{sec:Application}
we will then derive (!) equations (\ref{eqv4_08}) and (\ref{eqv4_12})
by means of the developed information transfer concept (see equations
(\ref{eqv3_13}) and (\ref{eqv3_28})).

The information transfer concept allows one to understand the occurrence
of equations (\ref{eqv4_08}) and (\ref{eqv4_12}) from a very fundamental
point of view which is based on information theory. However, this
concept can principally not explain why the constant factor is $1/2$
in equation (\ref{eqv4_08}) and $2/f$ in equation (\ref{eqv4_12}).
The explanation of these constants requires detailed knowledge or
correct assumptions, respectively, of the internal/physical mechanisms
of the observed process (see chapter \ref{sec:Phys-interpretation}).
This means that all equations derived by the generalized information
transfer concept \cite{Fielitz2011} are quantitative but phenomenological
in the description of natural processes (see section \ref{sub:Basics-Concluding}).

\subsubsection{Outlook to the next chapters}

In chapter \ref{sec:Basics} we repeat the basics of the generalized
concept of information transfer in order to update some important
notations introduced in reference \cite{Fielitz2011}, which improves
the clarity of the presentation of the concept. In appendix \ref{sec:Appendix-Comments}
we compile all comments related to the update of notations. For the
reader who wants to apply our approach to his own problems, it is,
however, highly recommended to study also reference \cite{Fielitz2011}
if he wants to understand all aspects (e.g. the derivation of theorems
\ref{theo:01} and \ref{theo:02} from three axioms) of the information
transfer concept.

In chapter \ref{sec:Application} we demonstrate the validity and
the applicability of the information transfer concept by three well
understood physical processes.

In chapter \ref{sec:Photon-propagation} we apply the information
transfer concept to the photon propagation process from a star. An
observational indication that not \emph{all aspects} of the photon
propagation process from a star are really well understood is the
so called K-Trumpler effect. The K-Trumpler effect was first noticed
in 1911 and is since then a challenging puzzle \cite{Arp1992}.

In chapter \ref{sec:Phys-interpretation} we demonstrate by two physical
examples that the natural amount of equilibrium information is not
an abstract number but has real physical meaning.

\section{Basics of the generalized information transfer concept\label{sec:Basics}}

Hartley \cite{Hartley1928} \emph{postulated} in 1928 that the transmitted
amount of information, $I$, is proportional to the number, $n$,
of selected symbols 
\begin{equation}
I=K\cdot n\label{eqv4_13}
\end{equation}
and for the postulated constant, $K$, Hartley concluded (the logarithmic
base is arbitrary) 
\begin{equation}
K_{{\rm Hartley}}=\frac{I}{n}=\ln s\label{eqv4_14}
\end{equation}
where $s$ is the number of available symbols at each selection. Equation
(\ref{eqv4_14}) is called Hartley function in the literature \cite{Klir1998}
and follows also from the Shannon entropy \cite{Shannon1948} 
\begin{equation}
K_{{\rm Shannon}}=\frac{I}{n}=-\sum\limits _{j=1}^{s}p_{j}\ln p_{j}\label{eqv4_15}
\end{equation}
if the probability to select a symbol is equal for all available symbols,
i.e. $p_{j}=1/s={\rm const.}$

The generalized information theory \cite{Klir1998} is concerned with
information conceived in terms of uncertainty reduction. Klir and
Wierman (\cite{Klir1998} p.~135) pointed out the differences between
the Hartley function (\ref{eqv4_14}) and the Shannon entropy (\ref{eqv4_15}):
``We recognize now that the Hartley function and the Shannon entropy
measure distinct types of uncertainty, nonspecificity and strife,
respectively. This distinction was concealed within the confines of
classical information theory, in which the Hartley function is almost
routinely viewed as a special case of the Shannon entropy, emerging
for uniform probability distributions. This view, which is likely
a result of the fact that the value of the Shannon entropy of the
uniform probability distribution on some set is equal to the value
of the Hartley function for the same set, was ill-conceived. Indeed,
the Hartley function is totally independent of any probabilistic assumptions,\ldots{}''

The fact that the Hartley function is not merely a special case of
the Shannon entropy is an important result of the generalized information
theory \cite{Klir1998} because it enables us to have nowadays an
unbiased view to Hartley's postulation (\ref{eqv4_13}) of an amount
of information. Natural processes can often be quantitatively described
by an absolute value of a process variable $z$, like e.g., a distance
value, a time interval value, a volume value, an energy value etc.
Hartley's postulation (\ref{eqv4_13}) of the amount of information
allows us to define a \emph{natural} amount of information, $I_{z}$,
to a generic process variable $z$ \cite{Fielitz2011} 
\begin{equation}
I_{z}=\kappa_{z}\frac{\left|\Delta z\right|}{\left|\delta z\right|}\quad{\rm with}\quad0<\kappa_{z}<\infty\label{eqv3_01}
\end{equation}
where $\left|\Delta z\right|$ is the \emph{absolute value} of the
process variable $z$, $\kappa_{z}$ is the \emph{information transfer
constant} of the process variable $z$, $\left|\delta z\right|$ is
the \emph{signal} of the process variable $z$ which has the property
that it is much smaller than the absolute value $\left|\Delta z\right|$,
that is $\left|\delta z\right|\ll\left|\Delta z\right|$. In appendix
\ref{sec:Appendix-Signals} it is explained why $\left|\delta z\right|$
is a signal and why the defined amount of information (equation (\ref{eqv3_01})),
which is related to a generic process variable, is a natural amount
of information.

Information theories \cite{Hartley1928}-\cite{Frieden1998} define
different measures of information related to finite sets (collections
of objects) and/or probability distributions, respectively. Methodological
minimum/maximum principles based on these measures are then applied.
The maximum entropy principle developed by Jaynes \cite{Jaynes1957}
is broadly utilized within classical information theory. However,
an extension of the maximum entropy principle to systems far from
thermodynamic equilibrium or even to non-physical systems is problematic
because it requires the choice of adequate constraints (see \cite{Haken2000}
for further discussion). The important point of the proposed generalized
concept of information transfer \cite{Fielitz2011} is that the defined
natural amount of information, equation (\ref{eqv3_01}), is not related
to finite sets and/or probability distributions, respectively. In
this way the application of a minimum/maximum principle becomes redundant,
and hence, the necessity to choose constraints which are adequate
to the considered system. Furthermore, if one goes back to the roots
of classical information theory \cite{Hartley1928}, \cite{Shannon1948}
it becomes obvious that the development of classical information theory
was strongly motivated by the request to transfer an amount of information
by technical processes. From the concrete systems (telegraphy, telephony,
television etc.) which realize technical information transfer processes
it could be abstracted in the classical information theory \cite{Hartley1928},
\cite{Shannon1948}. In analogy one can now abstract from the concrete
natural systems where information transfer processes occur.

With the definition of a natural amount of information (\ref{eqv3_01})
one can model information transfer from a generic process variable
$q$ (the information source) to a generic process variable $u$ (the
information destination) in a very abstract way. For that one must,
however, consider sufficiently simple natural processes which obey
the following condition.

\begin{condition}\label{cond:01} 

The considered natural process can be described by only two independent
generic process variables $(q,u)$ and is able to transfer information.
That is, the natural amount of information $I_{u}$ which the information
destination $u$ receives is not zero $\left(I_{u}>0\right)$.

\end{condition}
\begin{figure}
\includegraphics[width=7.5cm]{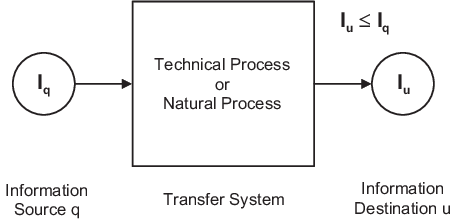} \protect\caption{Let the information source $q$ have the amount of information $I_{q}$.
Any process can transfer at best the complete amount of information
$I_{q}$ from an information source $q$ to an information destination
$u$ so that we have generally the inequality $I_{u}\le I_{q}$.}

\label{fig:01} 
\end{figure}

It is well known from Shannon's information theory \cite{Shannon1948}
that any technical process can transfer at best the complete amount
of information $I_{q}$ from an information source $q$ to an information
destination $u$ (Fig.~\ref{fig:01}). That is, one has always for
the amount of information, $I_{u}$, which the information destination
$u$ receives from the information source $q$, the inequality 
\begin{equation}
I_{u}\le I_{q}\label{eqv3_02}
\end{equation}
It is implicitly obvious from information theory \cite{Shannon1948}
that inequality~(\ref{eqv3_02}) is valid for any process so that
it can also be applied for simple natural processes which obey condition~\ref{cond:01}.
From inequality~(\ref{eqv3_02}) one gets a qualitative indication
that \emph{system independent correlations} should exist between two
generic variables of a simple natural process. The quantitative derivation
of such system independent correlations is described in detail in
reference \cite{Fielitz2011}. Here we will only explain the main
steps of this approach. First one has to introduce a further condition
so that it becomes possible to decide which process variable has the
status of an information source and which process variable has the
status of an information destination. This is possible if one considers
only natural processes which obey the following condition.

\begin{condition}\label{cond:02} The considered natural process
fulfills condition~\ref{cond:01}. One process variable, labeled
$x$, is a variable which has absolute values in the range $\left(0<\left|\Delta x\right|\le\infty\right)$.
The conjugate variable, labeled $y$, has absolute values in the range
$\left(0\le\left|\Delta y\right|<\infty\right)$.

\end{condition}

The generic process variables $\left(q,u\right)$ are related to processes
which obey condition~\ref{cond:01} whereas the generic process variables
$\left(y,x\right)$ are related to processes which obey condition~\ref{cond:02}.
We refer to variable $x$ as the (macroscopically) unrestricted process
variable and to variable $y$ as the (macroscopically) restricted
process variable. That is, we now consider only processes where one
variable is macroscopically restricted by specific conditions whereas
the conjugate variable is macroscopically unrestricted. As an example
consider the process of moving gas particles in a box. If the gas
box is located in a large heat reservoir at constant temperature the
amount of internal energy, $E\equiv\left|\Delta y\right|$, which
is related to the moving gas particles, is macroscopically restricted,
i.e. $E$ remains (in a real experiment) approximately constant if
the gas box volume increases (slowly). However, the gas box volume,
$V\equiv\left|\Delta x\right|$, is macroscopically an unrestricted
variable of this process, because one can experimentally choose a
gas box of any large size so that $V$ can (mathematically) have values
up to infinity.

For a simple process defined by condition~\ref{cond:02} one can
show \cite{Fielitz2011} that the restricted process variable $y$
has the status of an information source whereas the conjugate unrestricted
variable $x$ has the status of an information destination. In the
next step we define \emph{information equilibrium} (see also (a) in
appendix \ref{sec:Appendix-Comments})

\begin{definition}\label{def:01} \emph{\underline{I}nformation
\underline{E}quilibrium (IE)} is reached if the natural amount of
information, $I_{x}$, of process variable $x$ (defined by equation
(\ref{eqv3_01})) is equal to the natural amount of information, $I_{y}$,
of process variable $y$, that is, one has $I_{x}=I_{y}$.

\end{definition}and can then derive \cite{Fielitz2011} the \emph{global
IE theorem}

\begin{theorem}\label{theo:01} We consider any natural process which
fulfills condition~\ref{cond:02}. The process will reach information
equilibrium according to definition \ref{def:01} for a sufficiently
large absolute value $\left(\left|\Delta x\right|\to\infty\right)$
of the (macroscopically) unrestricted process variable $x$. If the
considered process has reached information equilibrium for any absolute
value $\left|\Delta x^{*}\right|$ of the process variable $x$, it
will stay in the state of information equilibrium for $\left|\Delta x\right|\ge\left|\Delta x^{*}\right|$.

\end{theorem}and the \emph{local IE theorem} (see also (b) in appendix
\ref{sec:Appendix-Comments})

\begin{theorem}\label{theo:02} We consider any natural process which
fulfills condition~\ref{cond:02}. The process will reach information
equilibrium according to definition \ref{def:01} for a sufficiently
small absolute value $\left(\left|\Delta y\right|\to0\right)$ of
the (macroscopically) restricted process variable $y$. If the considered
process has reached information equilibrium for any absolute value
$\left|\Delta y^{*}\right|$ of the process variable $y$, it will
stay in the state of information equilibrium for $\left|\Delta y\right|\le\left|\Delta y^{*}\right|$.

\end{theorem}

Considering equation (\ref{eqv3_01}) theorems \ref{theo:01} and
\ref{theo:02} can be expressed by the following equation 
\begin{equation}
\kappa_{x}\frac{\left|\Delta x\right|}{\left|\delta x\right|}=\kappa_{y}\frac{\left|\Delta y\right|}{\left|\delta y\right|}\quad{\rm if}\quad\left\{ \begin{array}{c}
\left|\Delta x\right|\ge\left|\Delta x^{*}\right|{\rm ,global}\\
\left|\Delta y\right|\le\left|\Delta y^{*}\right|{\rm ,local}
\end{array}\right.\label{eqv3_03}
\end{equation}
to which we refer as the \emph{IE equation} (see also (c) in appendix
\ref{sec:Appendix-Comments}). This equation (theorems \ref{theo:01}
and \ref{theo:02}) is system independent and will be correct in all
natural systems where one can observe simple processes which obey
condition~\ref{cond:02}. The problem is, however, that one often
cannot directly measure the signals $\left(\left|\delta x\right|,\left|\delta y\right|\right)$
of a natural process so that it is necessary to rearrange equation
(\ref{eqv3_03}) in this way 
\begin{equation}
\left(\frac{\left|\delta y\right|}{\left|\delta x\right|}\right)_{{\rm detector}}=\frac{\kappa_{y}}{\kappa_{x}}\frac{\left|\Delta y\right|}{\left|\Delta x\right|}\quad{\rm if}\quad\left\{ \begin{array}{c}
\left|\Delta x\right|\ge\left|\Delta x^{*}\right|{\rm ,global}\\
\left|\Delta y\right|\le\left|\Delta y^{*}\right|{\rm ,local}
\end{array}\right.\label{eqv3_04}
\end{equation}
to which we refer as the \emph{detector equation} and call the ratio
$\kappa_{y}/\kappa_{x}$ the \emph{information transfer constant ratio}.
The detector term in equation (\ref{eqv3_04}) is defined in the following
general way.

\begin{definition}\label{def:02} A \emph{detector} is an idealized
object with the ability to detect signals $\left(\left|\delta x\right|,\left|\delta y\right|\right)$
and to deliver as output value the ratio $\left(\left|\delta y\right|/\left|\delta x\right|\right)_{{\rm detector}}$.

\end{definition} 

The information transfer constant ratio $\kappa_{y}/\kappa_{x}$ in
equation (\ref{eqv3_04}) cannot be quantified in the context of the
information transfer model and must be determined experimentally or
by a more specific model according to the related physics (e.g. thermodynamics
and diffusion theory, as shown in the examples of chapter \ref{sec:Application}).
Furthermore, the absolute value $\left|\Delta x^{*}\right|$ or $\left|\Delta y^{*}\right|$,
respectively, is not quantitatively defined either (see derivation
of theorems \ref{theo:01} and \ref{theo:02} in the appendix of reference
\cite{Fielitz2011}).

\subsection{General solutions for simple processes which are in the state of
natural information equilibrium}

Theorems \ref{theo:01} and \ref{theo:02} formulate sufficient conditions
of information equilibrium for simple natural processes. However,
these conditions are not necessary conditions as will be demonstrated
in the discussion of the tracer diffusion process (see comments below
equation (\ref{eqv3_29})). The (sufficient) conditions for information
equilibrium are quite general so that one can expect that information
equilibrium will occur quite often in nature. In this section we now
presume a natural process which is in the state of information equilibrium
so that it is useful to define the corresponding process (this is
a newly introduced definition, since version v3 of this arXiv paper).

\begin{definition}\label{def:03} An \emph{IE process} is any natural
process which fulfills condition \ref{cond:02} and which is in the
state of information equilibrium (IE) according to definition \ref{def:01}.

\end{definition}

The detector equation (\ref{eqv3_04}) is valid for an IE process
so that one can apply differential calculus to the detector equation.
However, the absolute values of the signals $\left(\left|\delta x\right|,\left|\delta y\right|\right)$
must be small enough to be considered as infinitesimal increments
$\left(d\left|\Delta x\right|,d\left|\Delta y\right|\right)$ of the
absolute values $\left(\left|\Delta x\right|,\left|\Delta y\right|\right)$
of the process variables $\left(x,y\right)$. Before one integrates
the detector equation (\ref{eqv3_04}) one must distinguish strictly
between so called \emph{constant-restriction-parts} and \emph{floating
restrictions}.

\begin{definition}\label{def:04} A \emph{constant-restriction-part}
is given if the absolute value of the restricted process variable
$y$ is constant $\left(\left|\Delta y\right|=\left|\Delta y\right|_{{\rm const}}={\rm constant}\right)$
in the detector equation (\ref{eqv3_04}). If this is not the case
a \emph{floating restriction} is given. \end{definition}

For an IE process according to definition \ref{def:03} with a constant-restriction-part
according to definition \ref{def:04} the detector equation (\ref{eqv3_04})
becomes 
\begin{equation}
\frac{d\left|\Delta y\right|}{d\left|\Delta x\right|}=\pm\frac{\kappa_{y}}{\kappa_{x}}\frac{\left|\Delta y\right|_{{\rm const}}}{\left|\Delta x\right|}\label{eqv3_05}
\end{equation}
If we integrate equation (\ref{eqv3_05}) we have 
\begin{equation}
\frac{1}{\kappa_{x}}\int\limits _{\left|\Delta x\right|_{{\rm ref}}}^{\left\langle \left|\Delta x\right|\right\rangle }\frac{1}{\left|\Delta x\right|}\, d\left|\Delta x\right|=\pm\frac{1}{\kappa_{y}\left|\Delta y\right|_{{\rm const}}}\int\limits _{\left|\Delta y\right|_{{\rm ref}}}^{\left|\Delta y\right|}d\left|\Delta y\right|^{\prime}\label{eqv3_06}
\end{equation}
where $\left\langle \left|\Delta x\right|\right\rangle $ is the \emph{expected}
absolute value of the unrestricted process variable $x$ and $\left|\Delta y\right|$
is a \emph{given} absolute value of the restricted process variable
$y$. The subscript \emph{ref} indicates reference values of the IE
process. The absolute value $\left|\Delta y\right|_{{\rm {const}}}$
is the \emph{constant-restriction-part} of the considered process.
The general solution of equation (\ref{eqv3_06}) is 
\begin{equation}
\left\langle \left|\Delta x\right|\right\rangle =\left|\Delta x\right|_{{\rm ref}}\exp\left(\pm\frac{\kappa_{x}}{\kappa_{y}}\frac{\left|\Delta y\right|-\left|\Delta y\right|_{{\rm ref}}}{\left|\Delta y\right|_{{\rm const}}}\right)\label{eqv3_07}
\end{equation}
to which we refer as the \emph{constant-restriction-part solution}
of an IE process.

For an IE process with a floating restriction according to definition
\ref{def:04} the detector equation (\ref{eqv3_04}) becomes 
\begin{equation}
\frac{d\left|\Delta y\right|}{d\left|\Delta x\right|}=\pm\frac{\kappa_{y}}{\kappa_{x}}\frac{\left|\Delta y\right|}{\left|\Delta x\right|}\label{eqv3_08}
\end{equation}
If we integrate this equation we have 
\begin{equation}
\frac{1}{\kappa_{x}}\int\limits _{\left|\Delta x\right|_{{\rm ref}}}^{\left\langle \left|\Delta x\right|\right\rangle }\frac{1}{\left|\Delta x\right|}\, d\left|\Delta x\right|=\pm\frac{1}{\kappa_{y}}\int\limits _{\left|\Delta y\right|_{{\rm ref}}}^{\left|\Delta y\right|}\frac{1}{\left|\Delta y\right|^{\prime}}\, d\left|\Delta y\right|^{\prime}\label{eqv3_09}
\end{equation}
Solving the integrals gives the general solution 
\begin{equation}
\left\langle \left|\Delta x\right|\right\rangle ^{\frac{1}{\kappa_{x}}}=R_{\kappa}\left|\Delta y\right|^{\pm\frac{1}{\kappa_{y}}}\;{\rm with}\; R_{\kappa}=\frac{\left(\left|\Delta x\right|_{{\rm ref}}\right)^{\frac{1}{\kappa_{x}}}}{\left(\left|\Delta y\right|_{{\rm ref}}\right)^{\pm\frac{1}{\kappa_{y}}}}\label{eqv3_10}
\end{equation}
where $\left\langle \left|\Delta x\right|\right\rangle $ is the \emph{expected}
absolute value of the unrestricted process variable $x$ and $\left|\Delta y\right|$
is a \emph{given} absolute value of the restricted process variable
$y$. The subscript \emph{ref} indicates reference values of the IE
process. We refer to this equation as the \emph{floating solution}
of an IE process. It is convenient to call $R_{\kappa}$ the \emph{reference
constant} of an IE process.

We introduced a detector by definition \ref{def:02}. It is also useful
to define an \emph{unrestricted detector}.

\begin{definition}\label{def:05} An \emph{unrestricted detector}
is given if the absolute value $\left|\Delta x\right|$ of the unrestricted
process variable $x$ in the detector equation (\ref{eqv3_04}) is
always equal to the expected absolute value $\left\langle \left|\Delta x\right|\right\rangle $
of the unrestricted process variable $x$ $\left(\left|\Delta x\right|=\left\langle \left|\Delta x\right|\right\rangle \right)$.

\end{definition}

The special feature of an unrestricted detector according to definition
\ref{def:05} is that one can combine the detector equation (\ref{eqv3_04})
and the general solutions of an IE process (equation (\ref{eqv3_07})
or equation (\ref{eqv3_10}), respectively). We will see that this
will result also in well known laws (see equations (\ref{eqv3_21})
and (\ref{eqv3_22})).

\subsection{Concluding remarks\label{sub:Basics-Concluding}}

Because of the fundamental inequality (\ref{eqv3_02}) of information
theory one can assume that system independent laws should exist between
two independent generic variables of a simple process (simplicity
defined by condition \ref{cond:02}). The quantitative derivation
of such system independent laws (equations (\ref{eqv3_04}), (\ref{eqv3_07})
and (\ref{eqv3_10})) is described in more detail in \cite{Fielitz2011}.
In the context of the generalized information transfer concept it
now becomes understandable why \emph{exponential laws} (\ref{eqv3_07})
and \emph{power laws} (\ref{eqv3_10}) are often sufficient to describe
totally different simple processes in totally different systems. This
is an enormous generalization in the understanding of simple processes
in nature. From this point of view we now see the reasons why exponential
laws and power laws are often sufficiently \emph{quantitative} in
the description of simple systems/processes, but now we also see clearly
that these laws are \emph{phenomenological} as long as we do not understand
the meaning of the constant factors which comprise the intrinsic properties
of a specific system. That is, a deeper understanding of a specific
system in nature is directly related to the understanding of the constant
factors of the exponential laws and the power laws in nature.

The exponential law (\ref{eqv3_07}) and the power law (\ref{eqv3_10})
are integral representations of the \emph{detector equation} (\ref{eqv3_04}).
However, the detector equation (\ref{eqv3_04}) is more fundamental
than its integral representations because it contains only one single
dimensionless system dependent constant. This single dimensionless
constant is in fact a ratio, $\kappa_{y}/\kappa_{x}$, of two dimensionless
information transfer constants. The information transfer constant
ratio in the detector equation (\ref{eqv3_04}) cannot be quantified
in the context of the presented information transfer concept. This
means that the detector equation (\ref{eqv3_04}) is \emph{quantitative}
but \emph{phenomenological} in the description of natural processes
because the unknown constant ratio, $\kappa_{y}/\kappa_{x}$, must
be determined experimentally or from a more specific model of the
considered system -- if available.

\section{Application of the global IE theorem to well understood physical
processes\label{sec:Application}}

In this chapter we apply the global IE theorem \ref{theo:01} to three
well understood physical processes. In this way we can demonstrate
that this theorem is applicable and valid for simple processes which
obey condition \ref{cond:02}. In reference \cite{Fielitz2011} the
interested reader will also find an application of the local IE theorem
\ref{theo:02} which yields Fick's first well known law in the case
of the random walk process.

\subsection{The process of moving gas particles in a box}

To apply the detector equation (\ref{eqv3_04}) one must first define
an appropriate detector which corresponds to definition \ref{def:02}.
In the introduction (section \ref{sub:Pressure-Meas-Device}) we considered
a mechanical gas pressure measurement device 
\begin{equation}
p=\left(\frac{\left|F\right|}{A}\right)_{{\rm device}}\label{eqv3_11}
\end{equation}
where $\left|F\right|$ is the measured force which acts on a membrane
with the area $A$ and we concluded that $\left(\left|F\right|/A\right)_{{\rm device}}=\left|\partial E\right|/\left|\partial V\right|$
(see equation (\ref{eqv4_10})). In the context of the information
transfer concept one considers $\left|\partial E\right|/\left|\partial V\right|$
as a term which corresponds to definition \ref{def:02} of a detector
\begin{equation}
p=\left(\frac{\left|F\right|}{A}\right)_{{\rm device}}=\left(\frac{\left|\delta E\right|}{\left|\delta V\right|}\right)_{{\rm detector}}\label{eqv3_12}
\end{equation}
In the next step one must decide whether the process of moving gas
particles in a box is simple enough to obey condition \ref{cond:02}.
That is, whether energy and volume are generic and sufficient to describe
this process. This is true for ideal gas particles.

Before one applies the global IE theorem \ref{theo:01} one must first
know which generic process variable is restricted. If the gas box
is located in a large heat reservoir at constant temperature the amount
of internal energy, $E$, which is related to the moving gas particles,
is macroscopically restricted, i.e. $E$ remains (in a real experiment)
approximately constant if the gas box volume increases (slowly). According
to the global IE theorem \ref{theo:01} one can now conclude that
the considered process (moving gas particles in a box) must reach
information equilibrium if one enlarges the gas box volume sufficiently,
i.e. $V\ge V^{*}$. Applying the detector equation (\ref{eqv3_04})
(the global case) and considering equation (\ref{eqv3_12}) one gets
for the considered process 
\begin{equation}
p=\left(\frac{\left|F\right|}{A}\right)_{{\rm device}}=\left(\frac{\left|\delta E\right|}{\left|\delta V\right|}\right)_{{\rm detector}}=\frac{\kappa_{E}}{\kappa_{V}}\frac{E}{V}\quad{\rm if}\quad V\ge V^{*}\label{eqv3_13}
\end{equation}
Because the necessary (large) volume amount $V^{*}$ to reach information
equilibrium is not quantitatively defined in the context of the information
transfer model one can only say that a real gas will become an ideal
gas and obeys equation (\ref{eqv3_13}) if the gas volume, $V$, becomes
sufficiently large. This conclusion follows directly from the global
IE theorem \ref{theo:01} and is well known experimental experience
related to real gases. Furthermore, in the context of the information
transfer concept the information transfer constant ratio, $\kappa_{E}/\kappa_{V}$,
must be determined experimentally or from a more specific model. In
this case one can apply well known results from the thermodynamics
of the ideal gas 
\begin{equation}
p=\frac{2}{f}\frac{U}{V}\label{eqv3_14}
\end{equation}
where $U\equiv E$ is the internal energy of the ideal gas, $V$ the
ideal gas volume and $f$ the degree of freedom per ideal gas particle.
Comparing equations (\ref{eqv3_13}) and (\ref{eqv3_14}) one can
conclude for the information transfer constant ratio $\kappa_{E}/\kappa_{V}=2/f$
and one can now write the IE equation (\ref{eqv3_03}) in this way
\begin{equation}
I_{V}=f\frac{V}{\left|\delta V\right|}=2\frac{E}{\left|\delta E\right|}=I_{E}\quad{\rm if}\quad V\ge V^{*}\label{eqv3_15}
\end{equation}
That is, in the context of the information transfer concept we say
(in a very abstract but fundamental way) that the real gas reaches
an information equilibrium $\left(I_{V}=I_{E}\right)$ and becomes
an ideal gas if the gas volume becomes sufficiently large (for a given
number of gas particles). In the context of thermodynamics we say
that the real gas becomes an ideal gas if the average spacing between
the gas particles becomes large enough so that the gas particles can
be considered as an ensemble of point particles with no form of interaction.

In chapter \ref{sec:Phys-interpretation} we show that the natural
amount of information which is reached in information equilibrium
is given by $I_{V}=I_{E}=f\cdot N\gg1$ and is, hence, proportional
to the number of ideal gas particles and proportional to the degree
of freedom per gas particle.

\subsection{The fall process of a physical body in a gravitational field}

If one drops a small physical body on the surface of a planet the
physical body is accelerated by the gravitational force at the surface
and one can observe its velocity $v=dl/dt$. In the context of the
information transfer concept one can now consider the absolute value
of the velocity of the falling physical body (point mass) 
\begin{equation}
\left|v\right|=\left(\frac{\left|\delta l\right|}{\left|\delta t\right|}\right)_{{\rm detector}}\label{eqv3_16}
\end{equation}
which corresponds to the definition \ref{def:02} of a detector. That
is, we consider the falling physical body as a detector. Next one
must see which is the restricted variable of the fall process. This
is the length variable because one selects a fall height, $\left|\Delta l\right|$,
of limited length. However, the time interval $\left|\Delta t\right|$
the physical body will need to fall from the given fall height is
not a priori restricted. On planets or satellites with very low masses
it will require a very long fall time for a given fall height. That
is, according to the global IE theorem \ref{theo:01} the fall process
of a physical body on a planet must reach information equilibrium
if the fall time, $\left|\Delta t\right|$, becomes sufficiently large
so that one gets according to the detector equation (\ref{eqv3_04})
and equation (\ref{eqv3_16}) 
\begin{equation}
\left|v\right|=\left(\frac{\left|\delta l\right|}{\left|\delta t\right|}\right)_{{\rm detector}}=\frac{\kappa_{l}}{\kappa_{t}}\frac{\left|\Delta l\right|}{\left|\Delta t\right|}\quad{\rm if}\quad\left|\Delta t\right|\ge\left|\Delta t^{*}\right|\label{eqv3_17}
\end{equation}
The information transfer constant ratio, $\kappa_{l}/\kappa_{t}$,
must be determined experimentally or by a more specific model, respectively,
and the value of $\left|\Delta t^{*}\right|$ cannot be quantified
a priori either. Furthermore, it is required that the process is simple
enough (condition \ref{cond:02}), that is, the variables time and
length are generic and sufficient to describe the fall process of
a small physical body on the surface of a planet. This is only the
case if the fall process occurs in a sufficiently good vacuum. From
classical mechanics (constant acceleration of a point mass) one gets
\begin{equation}
\left|v\right|=\left(\frac{\left|\delta l\right|}{\left|\delta t\right|}\right)_{{\rm detector}}=2\frac{\left|\Delta l\right|}{\left|\Delta t\right|}\label{eqv3_18}
\end{equation}
and, hence, $\kappa_{l}/\kappa_{t}=2$. For an IE process according
to definition \ref{def:03} one can apply its floating solution (\ref{eqv3_10}),
with $\kappa_{l}/\kappa_{t}=2$ and a positive sign, and gets on the
surface of a planet (in vacuum) 
\begin{equation}
\left\langle \left|\Delta t\right|\right\rangle ^{2}=R_{{\rm Planet}}\left|\Delta l\right|\quad{\rm with}\quad R_{{\rm Planet}}=\frac{\left|\Delta t\right|_{{\rm ref}}^{2}}{\left|\Delta l\right|_{{\rm ref}}}\label{eqv3_19}
\end{equation}
where $\left\langle \left|\Delta t\right|\right\rangle $ is the \emph{expected}
time interval the physical body will need to fall from a \emph{given}
height, $\left|\Delta l\right|$, and $R_{{\rm Planet}}$ is the \emph{reference
constant} of this IE process (fall process on the surface of a planet).
This example demonstrates that the detector equation (\ref{eqv3_18})
is more fundamental than its integral representation (\ref{eqv3_19})
because there is no planet (mass) dependent constant in equation (\ref{eqv3_18}).
On the Earth's surface the reference constant, $R_{{\rm Earth}}=2/g$,
is given by the acceleration constant, $g$, which has been determined
experimentally so that one has 
\begin{equation}
\left\langle \left|\Delta t\right|\right\rangle ^{2}=\frac{2}{g}\left|\Delta l\right|\label{eqv3_20}
\end{equation}
The physical body (detector) is unrestricted according to definition
\ref{def:05} so that one has $\left|\Delta t\right|=\left\langle \left|\Delta t\right|\right\rangle $.
If a detector is unrestricted one can combine the detector equation
(\ref{eqv3_18}) and its floating solution (\ref{eqv3_20}). Doing
this one gets the well known length-independent equation 
\begin{equation}
\left|v\right|=g\left|\Delta t\right|\label{eqv3_21}
\end{equation}
and the well known time-independent equation on the Earth's surface
in vacuum 
\begin{equation}
\left|v\right|^{2}=2g\left|\Delta l\right|\label{eqv3_22}
\end{equation}
At first glance this approach to derive equations (\ref{eqv3_21})
and (\ref{eqv3_22}) looks like a trivial mathematical conversion:
It is, however, a fundamental feature of the information transfer
concept. This becomes obvious if one compares how the time-independent
equation (\ref{eqv3_22}) is derived in the usual way. For that one
considers the kinetic energy and potential energy of the physical
body (point mass) 
\begin{equation}
E_{{\rm kin}}=\frac{m_{{\rm inertial}}}{2}\left|v\right|^{2};\quad E_{{\rm pot}}=m_{{\rm heavy}}\, g\left|\Delta l\right|\label{eqv3_23}
\end{equation}
Let us recall in the context of equation (\ref{eqv3_23}) the long
lasting discussion in classical mechanics whether the inertial mass
and the heavy mass of a physical body are really identical. This question
was not decided before Einstein postulated the identity of inertial
and heavy mass. If one finally assumes that potential energy converts
in vacuum completely to kinetic energy one gets also equation (\ref{eqv3_22}).
The important (fundamental) feature of the derivation of equation
(\ref{eqv3_22}) via the information transfer concept is that one
does not need the physical concepts of mass and energy.

One could argue that one can simply start from the condition of constant
acceleration 
\begin{equation}
\frac{d^{2}l}{dt^{2}}=g={\rm const.}\label{eqv4_39}
\end{equation}
and can then also derive equation (\ref{eqv3_22}) mathematically
without making use of the concepts of mass and energy. However, the
concept behind equation (\ref{eqv4_39}) is an equilibrium of forces,
$F_{{\rm eq}}$, 
\begin{equation}
F_{{\rm eq}}=m_{{\rm inertial}}\frac{d^{2}l}{dt^{2}}=m_{{\rm heavy}}\, g\label{eqv4_40}
\end{equation}
so that equation (\ref{eqv4_39}) implies the assumption $m_{{\rm inertial}}=m_{{\rm heavy}}$,
and hence, the problem discussed above. In contrast the concept behind
equation (\ref{eqv3_17}) is an equilibrium of natural information,
$I_{{\rm eq}}$, 
\begin{equation}
I_{{\rm eq}}=\kappa_{t}\frac{\left|\Delta t\right|}{\left|\delta t\right|}=\kappa_{l}\frac{\left|\Delta l\right|}{\left|\delta l\right|}\label{eqv4_41}
\end{equation}
as defined by equation (\ref{eqv3_01}) which does not need the physical
concept of mass.

\subsection{The tracer particle propagation process in a solid body\label{sub:Tracer-particles}}

\begin{figure}
\includegraphics[width=7cm]{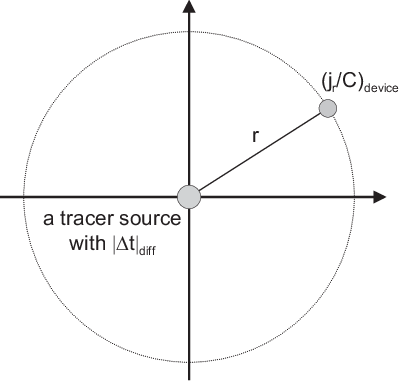} \protect\caption{In the center of the coordinate system a tracer particle source is
located with a given diffusion time interval $\left|\Delta t\right|_{{\rm diff}}$.
At the distance $r$ a virtual measurement device $\left(j_{r}/C\right)_{{\rm device}}$
delivers as output the local ratio between radial flux $j_{r}$ and
concentration $C$.}

\label{fig:02} 
\end{figure}

We now consider the diffusion process of tracer particles from a small
particle source located at $r=0$ (see Fig.~\ref{fig:02}). To study
the tracer particle diffusion process one defines a local radial tracer
particle flux 
\begin{equation}
j_{r}=\frac{1}{A_{0}}\frac{\partial N}{\partial t}\label{eqv3_24}
\end{equation}
where $\partial N$ is the infinitesimal number of tracer particles
which pass radially through the unit area $A_{0}$ during the infinitesimal
time interval $\partial t$. The local concentration of the tracer
particles is also given by definition as 
\begin{equation}
C=\frac{1}{A_{0}}\frac{\partial N}{\partial r}\label{eqv3_25}
\end{equation}
We can now define a local measurement device 
\begin{equation}
\left(\frac{j_{r}}{C}\right)_{{\rm device}}\label{eqv3_26}
\end{equation}
which yields the local ratio between flux and concentration at any
distance, $r$, from the tracer particle source (see Fig.~\ref{fig:02}).
It is a virtual measurement device because there is no technical realization
(at least to date) of such a measurement device in a solid medium.
Nevertheless, we can make thought experiments with this defined virtual
device. If we apply the definitions for the local radial flux (\ref{eqv3_24})
and the local concentration (\ref{eqv3_25}) we get, and this is an
important point, 
\begin{equation}
\left(\frac{j_{r}}{C}\right)_{{\rm device}}=\left(\frac{\left|\delta r\right|}{\left|\delta t\right|}\right)_{{\rm detector}}\label{eqv3_27}
\end{equation}
so that the defined virtual device corresponds to definition \ref{def:02}
of a detector.

To obey condition \ref{cond:02} the tracer particle diffusion process
must become sufficiently simple so that time and length are the only
generic variables of this process. To achieve this condition the medium
where diffusion occurs must be sufficiently simple which is the case
if we consider tracer diffusion in a sufficiently large homogeneous
solid medium. Next one must ask which process variable is restricted.
The time variable is restricted if one considers a diffusion source
with a given long diffusion time interval, $\left|\Delta t\right|_{{\rm diff}}$,
so that one can consider it approximately constant if one makes measurements
with the device $\left(j_{r}/C\right)_{{\rm device}}$ at time intervals
much shorter than the diffusion time interval. However, the length
variable is an unrestricted process variable if the tracer diffusion
process takes place in a sufficiently large volume of the medium so
that one can make measurements with the device $\left(j_{r}/C\right)_{{\rm device}}$
at sufficiently long distances $r$ from the tracer particle source
(see Fig.~\ref{fig:02}). According to the global IE theorem \ref{theo:01}
one can then conclude that the tracer diffusion process must reach
information equilibrium if the length variable (distance $r$) becomes
sufficiently large so that one gets according to the detector equation
(\ref{eqv3_04}) and equation (\ref{eqv3_27}) 
\begin{equation}
\left(\frac{j_{r}}{C}\right)_{{\rm device}}=\left(\frac{\left|\delta r\right|}{\left|\delta t\right|}\right)_{{\rm detector}}=\frac{\kappa_{r}}{\kappa_{t}}\frac{r}{\left|\Delta t\right|_{{\rm diff}}}\quad{\rm if}\quad r\ge r^{*}\label{eqv3_28}
\end{equation}
Note that equation (\ref{eqv4_08}) considered in the introduction
can be derived by the information transfer concept in the same way
(replace $r$ by $x$ and consider a one-dimensional diffusion process).
The undefined information transfer constant ratio, $\kappa_{r}/\kappa_{t}$,
must again be determined experimentally or by a more specific model.
In this case one can apply the diffusion theory \cite{Crank1975}
in solids and gets for an infinitesimally small tracer particle point
source in an infinite homogeneous medium (see appendix \ref{sec:Appendix-Derivation})
\begin{equation}
\left(\frac{j_{r}}{C}\right)_{{\rm device}}=\left(\frac{\left|\delta r\right|}{\left|\delta t\right|}\right)_{{\rm detector}}=\frac{1}{2}\frac{r}{\left|\Delta t\right|_{{\rm diff}}}\label{eqv3_29}
\end{equation}
and, hence, $\kappa_{r}/\kappa_{t}=1/2$. Equation (\ref{eqv3_29})
is even correct at an infinitesimally small distance $r$ from the
tracer particle source so that the diffusion process from an infinitesimally
small tracer particle point source in an infinite homogeneous medium
is clearly in the state of information equilibrium according to definition
\ref{def:01}. This example demonstrates that \emph{the global IE
theorem \ref{theo:01} formulates sufficient but not necessary conditions
for information equilibrium}.

Until now we have considered only the case where the local measurement
device $\left(j_{r}/C\right)_{{\rm device}}$ is located at a constant
distance $\left(dr/dt=0\right)$ from the tracer point source (see
Fig.~\ref{fig:02}). We will now assume that the distance, $r$,
from the diffusion source is sufficiently small and that the diffusion
time interval, $\left|\Delta t\right|_{{\rm diff}}$, is sufficiently
long so that according to equation (\ref{eqv3_29}) the output value
of the measurement device becomes practically zero. In this case the
flux, and hence, the concentration gradient of the tracer particles
is practically zero and is negligible. If we then start to move the
measurement device (a small distance), tracer particles will pass,
because of the movement of the device, through the unit area $A_{0}$
(see equation (\ref{eqv3_24})) of the device so that its output value
is practically given by the relative motion of the device, expressed
as $\left(dr/dt\right)_{{\rm kinematic}}$ 
\begin{equation}
\left(\frac{j_{r}}{C}\right)_{{\rm device}}=-\left(\frac{dr}{dt}\right)_{{\rm kinematic}}\quad{\rm if}\quad\left(\frac{r}{\left|\Delta t\right|_{{\rm diff}}}\right)\to0\label{eqv3_30}
\end{equation}
If the detector moves away from the diffusion source, i.e. $dr/dt>0$,
the output value of the measurement device becomes negative. If it
moves, however, towards the diffusion source, i.e. $dr/dt<0$, its
output value becomes positive. Including the independent kinematic
term (\ref{eqv3_30}) by superposition into equation (\ref{eqv3_29})
one gets 
\begin{equation}
\left(\frac{j_{r}}{C}\right)_{{\rm device}}=\left(\frac{1}{2}\frac{r}{\left|\Delta t\right|_{{\rm diff}}}\right)_{{\rm info.\; equi.}}-\left(\frac{dr}{dt}\right)_{{\rm kinematic}}\label{eqv3_31}
\end{equation}
so that the output value of the measurement device is given by an
information equilibrium term and by a kinematic term. Assuming we
would not be aware of the information equilibrium term in equation
(\ref{eqv3_31}) we would get the (wrong) impression from the measurement
device output that a tracer diffusion point source is moving \emph{towards}
the measurement device if the information equilibrium term becomes
dominant in equation (\ref{eqv3_31}). Interestingly enough we will
get a very similar equation if we discuss in the next chapter the
photon propagation process from a star (compare equations (\ref{eqv3_31})
and (\ref{eqv3_38})). However, in that case we will get the (wrong)
impression that a point source (a star) is moving \emph{away} from
the measurement device if the information equilibrium term becomes
dominant in equation (\ref{eqv3_38}).

\section{The photon propagation process from a star in vacuum\label{sec:Photon-propagation}}

As a non-trivial application of the information transfer concept \cite{Fielitz2011}
we will now discuss the photon propagation process from a star in
vacuum in this context. Stars form from dense cloud fragments of the
interstellar medium which is dominated by hydrogen and helium. Jones
\cite{Jones2007} describes the development of a star in this way:
``Detailed calculations show that the central regions of the fragment
contract the most rapidly. It is therefore these central regions that
become opaque to the photons emitted by the molecules. The temperature
rise is then rapid and the central object is regarded as a protostar.
Contraction continues, now more slowly, and a few million years after
the fragment separated from the dense cloud the temperature in the
core of the protostar has become high enough for nuclear fusion to
occur -- about $10^{7}$ K. This fusion releases energy and creates
a pressure gradient that halts the contraction of the protostar. At
this point the protostar has become a star -- a compact body sustained
by nuclear fusion.'' That is, if we consider the time $t=t_{0}$
where nuclear fusion occurs in the core of the protostar as the time
of birth of the star we have a defined starting time of the photon
propagation process. We observe the star now at $t=t_{1}$ so that
the time interval $\left|\Delta t\right|_{{\rm age}}=t_{1}-t_{0}$
becomes a generic time interval of the observed process. In this paper
the time of birth of the star is defined by the start of nuclear fusion
in the core of the protostar.

\begin{figure}
\includegraphics[width=7cm]{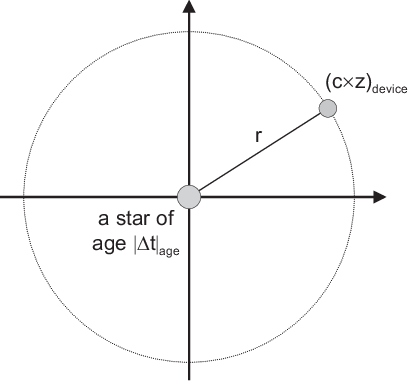} \protect\caption{In the center of the coordinate system a star is located which has
a given age, $\left|\Delta t\right|_{{\rm age}}$, with the time of
birth of the star being defined by the start of nuclear fusion in
the core of the protostar. At the distance $r$ a measurement device,
$\left(c\times z\right)_{{\rm device}}$, allows one a spectral analysis
of the light emitted by the star.}

\label{fig:03} 
\end{figure}

In Fig.~\ref{fig:03} we consider a star of a given age, $\left|\Delta t\right|_{{\rm age}}$.
At the distance $r$ we analyze the electromagnetic spectrum of the
star by a spectroscopic device, symbolized by $\left(c\times z\right)_{{\rm device}}$,
which delivers as output the redshift parameter $z$ where $c$ is
the velocity of light in vacuum. Redshift occurs when the electromagnetic
radiation that is emitted from a star is shifted towards the red end
(less energetic end or longer wavelength) of the electromagnetic spectrum.
The redshift parameter $z>0$ is used to describe the change in wavelength.
It is defined as \cite{Carroll2007} 
\begin{equation}
z=\frac{\lambda_{o}-\lambda_{e}}{\lambda_{e}}\label{eqv3_32}
\end{equation}
where $\lambda_{e}$ is the emitted and $\lambda_{o}$ is the observed
wavelength. Conversely, a decrease in wavelength is called blueshift
$(z<0)$. To exclude cosmological redshift we restrict the discussion
in this paper to stars of our own Milky Way Galaxy and to distances
in the order of the diameter of the Milky Way Galaxy, because the
Hubble expansion is not expected inside a galaxy.

It is well known experimental experience that a spectroscopic device
allows one to measure the radial velocity between the star and the
device so that one has \cite{Carroll2007} (Doppler effect)

\begin{equation}
\left(c\times z\right)_{{\rm device}}=\left(\frac{dr}{dt}\right)_{{\rm kinematic}}\quad{\rm if}\quad\left(\frac{dr}{dt}\right)_{{\rm kinematic}}\ll c\label{eqv3_33}
\end{equation}
If the device moves away from the star, i.e. $dr/dt>0$, the $z$
parameter becomes positive. If it moves towards the star, i.e. $dr/dt<0$,
the $z$ parameter becomes negative.

In the context of the information transfer model we now consider a
spectroscopic device at a constant distance $r$ (see Fig.~\ref{fig:03})
so that we have $\left(dr/dt\right)_{{\rm kinematic}}=0$. In this
way we can neglect any kinematic process (Doppler effect) and can
focus our attention to the photon propagation process from the star.
If a spectroscopic device delivers in this experimental situation
a redshift value 
\begin{equation}
\left(c\times z\right)_{{\rm device}}=\left(\frac{\left|\delta r\right|}{\left|\delta t\right|}\right)_{{\rm detector}}\quad{\rm if}\quad\left(\frac{dr}{dt}\right)_{{\rm kinematic}}=0\label{eqv3_34}
\end{equation}
it will correspond to definition \ref{def:02} of a detector so that
one can apply the detector equation (\ref{eqv3_04}) in the case of
information equilibrium. To be sure that time and distance are sufficient
and generic variables to describe the photon propagation process one
must consider a star at sufficiently long distances. In this case
the star becomes a point source where one generic variable of the
photon propagation process is the distance $r$ to this point. A second
generic variable is the age $\left|\Delta t\right|_{{\rm age}}$ of
the observed photon propagation process. Furthermore, the application
of the global IE theorem \ref{theo:01} requires a sufficiently simple
process (condition \ref{cond:02}). To achieve this condition one
must consider a single star in a sufficiently good vacuum (see also
the discussion related to equations (\ref{eqv3_17}) and (\ref{eqv3_27})).
That is, one can neglect the effect of cosmic gas or cosmic dust on
the propagating photons in the line of sight to the star.

For the application of the global IE theorem \ref{theo:01} one must
also decide which process variable is restricted and which is unrestricted.
Because one observes the photon propagation process always in a very
small time interval compared to the ages of the stars one can consider
the ages of the observed stars as given constants. That is, in the
context of the information transfer model the age $\left|\Delta t\right|_{{\rm age}}$
of a star is a restricted process variable. However, the distance
$r$ is an unrestricted process variable because one can move the
spectroscopic device $\left(c\times z\right)_{{\rm device}}$ in Fig.~\ref{fig:03}
in a thought experiment far enough away from the star so that one
must reach information equilibrium according to the global IE theorem
\ref{theo:01}. In reality one can simply choose any star at a sufficiently
large distance from the Earth to observe a photon propagation process
which is in the state of information equilibrium. One then has according
to the detector equation (\ref{eqv3_04}) and equation (\ref{eqv3_34})
$\left(\left(dr/dt\right)_{{\rm kinematic}}=0\right)$ 
\begin{equation}
\left(c\times z\right)_{{\rm device}}=\frac{\kappa_{r}}{\kappa_{t}}\frac{r}{\left|\Delta t\right|_{{\rm age}}}\quad{\rm if}\quad r\ge r^{*}\label{eqv3_35}
\end{equation}
We now see that the excess redshift (K-Trumpler effect) of a star
is proportional to the distance, $r$, and inversely proportional
to the age of the star, $\left|\Delta t\right|_{{\rm age}}$, with
the time of birth of the star being defined by the start of nuclear
fusion in the core of the protostar. The value $r^{*}$ is not quantitatively
defined in the context of the information transfer model. However,
one has to keep in mind that the global IE theorem \ref{theo:01}
formulates only sufficient conditions to reach information equilibrium.
We have seen that in the case of propagating tracer particles from
an infinitesimal tracer point source the particle propagation process
is completely in the state of information equilibrium (see discussion
related to equation (\ref{eqv3_29})). One can, therefore, assume
that equation (\ref{eqv3_35}) will be practically sufficiently accurate
if the star can be practically approximated as a point source. Near
the star the K-Trumpler effect becomes negligible because of the superposition
with the gravitational redshift and with mass-loss corrections \cite{Arp1992}.

As already explained in the examples discussed above the undefined
information transfer ratio, $\kappa_{r}/\kappa_{t}$, in equation
(\ref{eqv3_35}) must be determined by a more specific model or by
experimental observations, respectively. In this case there is no
more specific model available so that the constant ratio $\kappa_{r}/\kappa_{t}$
must be determined by astronomical observations. Considering the examples
discussed in chapter \ref{sec:Application} one can estimate an expected
range for the constant ratio of at least $0.5\le\kappa_{r}/\kappa_{t}\le2$.

To get a quantitative impression of the redshift value related to
information equilibrium we can set our sun in the center of Fig.~\ref{fig:03}
$\left(\left|\Delta t\right|_{{\rm age}}^{{\rm sun}}\cong4.6\times10^{9}yr\right)$
and can then observe our sun in a thought experiment at a distance
$r=10^{5}\, ly$ which is roughly the diameter of the Milky Way Galaxy
\cite{Carroll2007} 
\begin{equation}
\left(c\times z\right)_{{\rm device}}^{{\rm sun}}=\frac{\kappa_{r}}{\kappa_{t}}\frac{r}{\left|\Delta t\right|_{{\rm age}}^{{\rm sun}}}\cong\frac{\kappa_{r}}{\kappa_{t}}\frac{10^{5}ly}{4.6\times10^{9}yr}\cong\frac{\kappa_{r}}{\kappa_{t}}6.5\frac{km}{s}\label{eqv3_36}
\end{equation}
If one observes a star which is 10 times younger than the sun the
measured redshift value would become 10 times larger. (One would get
the same redshift value if one observes a star which is 10 times younger
than the sun at 1/10 of the distance.) We can compare the redshift
value (\ref{eqv3_36}) of the sun, which is related to information
equilibrium, with its gravitational redshift value \cite{Carroll2007}
\begin{equation}
\left(c\times z\right)_{{\rm device}}^{{\rm sun}}\cong\left(0.64\frac{km}{s}\right)_{{\rm gravitational}}\label{eqv3_37}
\end{equation}
and we see that it is about a factor of ten smaller and practically
negligible at the considered large distance $\left(r\cong10^{5}\, ly\right)$.

This discussion shows that the $z$ parameter is the result of two
superimposed independent processes: a photon propagation process in
the state of information equilibrium and a kinematic process (Doppler
effect) so that one finally gets 
\begin{equation}
\left(c\times z\right)_{{\rm device}}=\left(\frac{\kappa_{r}}{\kappa_{t}}\frac{r}{\left|\Delta t\right|_{{\rm age}}}\right)_{{\rm {\rm info.\; equi.}}}+\left(\frac{dr}{dt}\right)_{{\rm kinematic}}\label{eqv3_38}
\end{equation}
In equation (\ref{eqv3_38}) we neglected all effects which are related
to the properties/processes of the observed star like the gravitational
redshift and mass-loss corrections \cite{Arp1992}. Such constant
(distance independent) effects of the stars and also the Doppler effect
can be best studied on stars which are close to the Earth or very
old stars so that the information equilibrium term in equation (\ref{eqv3_38})
can be neglected. However, if one finally observes young stars which
are far away from the Earth and one is not aware of the information
equilibrium term in equation (\ref{eqv3_38}) one will get the (wrong)
impression that especially young stars are moving away from the observation
point (the Earth or solar system, respectively). This is indeed observed
and is called K effect or Trumpler effect, respectively. H. Arp \cite{Arp1992}
reviewed this effect in 1992 and writes in the introduction: ``The
first spectroscopic measurements of large numbers of B stars showed
that, unlike cooler stars, they appeared to be expanding away from
the solar neighbourhood. This positive redshift was expressed as a
'K term' and is referred to in the literature as the K effect. No
satisfactory explanation was ever advanced as to why the entire system
of luminous young stars should be receding from the position of the
Earth. When I was taking undergraduate courses in galactic dynamics
from Bart Bok in 1949 it was considered a mysterious and challenging
puzzle (\ldots{}). The same effect was reported by Trumpler who showed
that the brightest, hottest stars in young clusters had redshifts
which were systematically positive with respect to their clusters.
He obtained a mean excess of +10 km/s. Initially considerable interest
was generated by this result because it was thought to be a demonstration
of a gravitational redshift as predicted by Einstein's theory of general
relativity. But soon estimates of the surface gravity of such stars
made it clear that general relativity would predict only 1-2 km/s
shifts for the masses and radii which these stars possessed (\ldots{}).''

\subsection{Direct application of the information transfer concept\label{sub:Direct-application}}

Because of the system/process independent character of the global
IE theorem \ref{theo:01} it was useful to discuss first a simple
natural process which is basically similar (in the context of the
information transfer concept) to the photon propagation process from
a star in vacuum but which is, on the contrary, well understood. Such
a basically similar process is the three dimensional tracer particle
propagation process from a tracer particle point source in a homogeneous
medium (see section \ref{sub:Tracer-particles}). The analogy is that
photons are propagating from a point source and that tracer particles
are propagating (diffusing) from a point source. We saw that the application
of the detector equation (\ref{eqv3_04}) delivered very similar equations
for both discussed processes (compare equations (\ref{eqv3_31}) and
(\ref{eqv3_38})). The drawback is, however, that the application
of the detector equation (\ref{eqv3_04}) is very abstract. As explained
in chapter \ref{sec:Basics} (see discussion related to the IE equation
(\ref{eqv3_03})) the detector equation was introduced because often
one cannot directly measure the signals of a natural process. However,
if we discuss the photon propagation process from a star the experimental
situation is different because in this case \emph{signals} of the
generic process variables length and time can easily be measured.
This special situation allows us to derive equation (\ref{eqv3_35})
also in a more direct way by considering \emph{measured} natural amounts
of information. In this way the derivation of equation (\ref{eqv3_35})
becomes more descriptive. In the following discussion we do not speculate
about any internal/physical light propagation mechanisms which could
rationalize equation (\ref{eqv3_35}) but apply only facts.
\begin{description}
\item [{Fact}] \textbf{1}: An observed star has a given age, $\left|\Delta t\right|_{{\rm age}}$,
and one can identify absorption/emission lines in the measured spectrum
of the star, e.g. hydrogen absorption lines, with a known (reference)
frequency, $\nu_{{\rm ref}}$ \cite{Carroll2007}.
\end{description}
In the context of the information transfer concept any identified
reference frequency, $\nu_{{\rm ref}}$, in the measured spectrum
of the star is considered as a \emph{measured time signal}, $\left|\delta t\right|_{{\rm meas}}=1/\nu_{{\rm ref}}$.
Applying equation (\ref{eqv3_01}) one can evaluate a natural amount
of information related to the time variable 
\begin{equation}
I_{t}^{{\rm source}}=\kappa_{t}\,\frac{\left|\Delta t\right|_{{\rm age}}}{\left|\delta t\right|_{{\rm meas}}}=\kappa_{t}\,\left|\Delta t\right|_{{\rm age}}\nu_{{\rm ref}}={\rm const.}\label{eqv4_57}
\end{equation}
where $\left|\Delta t\right|_{{\rm age}}$ is the age of the star
with the time of birth of the star being defined (in this paper) by
the start of nuclear fusion in the core of the protostar. Because
astronomers can today estimate the age of an observed star \cite{Jones2007},
\cite{Carroll2007} the only unknown constant in equation (57) is
the information transfer constant, $\kappa_{t}$, related to the time
variable. Note that this natural amount of information is \emph{restricted}
(practically constant) during our observation time so that one can
identify this natural amount of information as the source of information
\cite{Fielitz2011}. This corresponds to our intuition that an observed
star should have the status of an information source.

In the next step we move in a thought experiment the spectroscopic
device to a very large distance (a distance in the order of the diameter
of the Milky Way Galaxy) away from the observed star and keep the
radial distance constant so that the Doppler effect is negligible.
We now make use of the following fact.
\begin{description}
\item [{Fact}] \textbf{2}: There is an excess redshift (K-Trumpler effect)
observed which cannot be interpreted by the Doppler effect \cite{Arp1992}.
\end{description}
If there is an excess redshift (K-Trumpler effect) one measures by
means of the spectroscopic device a length shift, $\left|\delta l\right|_{{\rm meas}}^{{\rm excess}}=\lambda_{{\rm obs}}-\lambda_{{\rm ref}}$,
of any identified reference wavelength of the observed star (which
cannot be interpreted by the Doppler effect). In the context of the
information transfer concept this small length shift is considered
as a \emph{measured length signal}. We assume in our thought experiment
that we know the distance, $r$, from the star so that we can evaluate
(if the constant $\kappa_{r}$ were known) a natural amount of information
related to the length/distance variable (applying equation (\ref{eqv3_01})))
\begin{equation}
I_{r}^{{\rm dest}}=\kappa_{r}\,\frac{r}{\left|\delta l\right|_{{\rm meas}}^{{\rm excess}}}=\kappa_{r}\frac{r}{\lambda_{{\rm obs}}-\lambda_{{\rm ref}}}\label{eqv4_58}
\end{equation}
Note that this natural amount of information has the status of an
information destination because the distance, $r$, is an \emph{unrestricted}
process variable \cite{Fielitz2011}. This corresponds to our intuition
that an observer of the star (on Earth) at the distance $r$ should
be in the status of a receiver of information. Finally we apply the
following fact from information theory (see discussion related to
inequality (\ref{eqv3_02}))
\begin{description}
\item [{Fact}] \textbf{3}: For the amount of information, $I_{r}^{{\rm dest}}$,
which the information destination receives (the Earth in this case)
from the information source (a star in this case), the inequality
$I_{r}^{{\rm dest}}\le I_{t}^{{\rm source}}$ is valid.
\end{description}
Every observed star sends an individual natural amount of information
(quantified by equation (\ref{eqv4_57})) to Earth. Fact 3 states
that one can principally \emph{not measure a larger} natural amount
of information (quantified by equation (\ref{eqv4_58})) on Earth
than the observed star sends. In the ideal case one can measure on
Earth at best the same natural amount of information, $I_{r}^{{\rm dest}}=I_{t}^{{\rm source}}$,
and would then reach information equilibrium (definition \ref{def:01}).
If one has reached information equilibrium for any distance $r^{*}$
the process remains in information equilibrium for larger distances
(see the global IE theorem \ref{theo:01}). Combining equations (\ref{eqv4_57})
and (\ref{eqv4_58}) one has (if $r\ge r^{*}$) 
\begin{equation}
I_{{\rm eq}}=\kappa_{r}\frac{r}{\lambda_{{\rm obs}}-\lambda_{{\rm ref}}}=\kappa_{t}\,\left|\Delta t\right|_{{\rm age}}\nu_{{\rm ref}}={\rm const.}\label{eqv4_59}
\end{equation}
where $I_{{\rm eq}}$ is the natural amount of equilibrium information.
We now see clearly that the measured excess wavelength shift, $\left|\delta l\right|_{{\rm meas}}^{{\rm excess}}=\lambda_{{\rm obs}}-\lambda_{{\rm ref}}$,
on Earth is proportional to the distance, $r$, because the sent (individual)
natural amount of information from the observed star is practically
constant during our observation time. 

Next one can apply the experimentally verified relation between wavelength,
$\lambda$, and frequency, $\nu$, in vacuum, $c=\lambda\cdot\nu$,
so that equation (\ref{eqv4_59}) becomes 
\begin{equation}
\frac{1}{c}\frac{\lambda_{{\rm ref}}}{\lambda_{{\rm obs}}-\lambda_{{\rm ref}}}=\frac{\kappa_{t}}{\kappa_{r}}\,\frac{\left|\Delta t\right|_{{\rm age}}}{r}\quad{\rm if}\quad r\ge r^{*}\label{eqv4_60}
\end{equation}
Considering the defined redshift parameter (equation (\ref{eqv3_32}))
one gets again equation (\ref{eqv3_35}) 
\begin{equation}
c\cdot z=\frac{\kappa_{r}}{\kappa_{t}}\,\frac{r}{\left|\Delta t\right|_{{\rm age}}}\quad{\rm if}\quad r\ge r^{*}\label{eqv4_61}
\end{equation}
The (alternative) derivation of equation (\ref{eqv4_61}) avoided
the introduction of an abstract ``device'' and an abstract ``detector''
but discussed directly \emph{measured} natural amounts of information.
This was possible because in this example of natural information transfer
the signals of the considered natural process can be easily measured.

\subsection{Concluding remarks}

If one can measure a signal (an excess wavelength shift, $\left|\delta l\right|_{{\rm meas}}^{{\rm excess}}=\lambda_{{\rm obs}}-\lambda_{{\rm ref}}$)
but cannot explain its physical/internal meaning it clearly indicates
that not yet \emph{all aspects} of the light propagation process on
large scales in vacuum are really well understood. Nevertheless, equation
(\ref{eqv4_61}) is quantitative and will be helpful for a quantitative
tracking of the K-Trumpler effect. If the K-Trumpler effect is confirmed
more quantitatively by astronomical observations it will deliver new
impulses to improve the theory of light propagation in vacuum on \emph{large
scales} (scales in the order of the diameter of a galaxy and larger)
because the Hubble expansion is not expected inside a galaxy.

The tendency that \emph{highly luminous stars} show an excess redshift
was first noticed in 1911 and called K effect, later, it was also
called Trumpler effect. Arp \cite{Arp1992} pointed to the qualitative
observation that the K-Trumpler effect is correlated to the age of
the stars:

Page 800 left column \cite{Arp1992}: ``No satisfactory explanation
was ever advanced as to why the entire system of luminous young stars
should be receding from the position of the Earth.''

Page 804 left column \cite{Arp1992}: ``The luminosities of these
supergiants, however, are also directly correlated with their evolutionary
ages in that more luminous isochrones represent younger stars.''

Page 809 right column \cite{Arp1992}: ``It is as if the whole isochrone
in the Herzsprung-Russel diagram has this excess redshift, and older
isochrones have less and less until we encounter normal-luminosity
stars.''

That is, the expected dependence of the K-Trumpler effect (equation
(\ref{eqv3_35}) or (\ref{eqv4_61}), respectively) on the age of
the stars is, at least qualitatively, already observed. The expected
distance dependence of the K-Trumpler effect of single stars has,
to our knowledge, not yet been reported. We assume the reason is that
the K-Trumpler effect is quite difficult to quantify correctly. According
to the discussion in this paper one can expect for the Sun an excess
redshift value of about 6.5 km/s (equation (\ref{eqv3_36}) if one
assumes $\kappa_{r}/\kappa_{t}\approx1$) if one observes the Sun
at a distance $r=10^{5}\, ly$ which is roughly the diameter of the
Milky Way Galaxy. However, the rotation curves \cite{Carroll2007}
of the Milky Way Galaxy show that the rotation velocities are in the
range 200 -- 250 km/s. This demonstrates the difficulties to separate
the K-Trumpler effect correctly from the Doppler effect inside the
Milky Way Galaxy.

Considering all previously discussed examples of information equilibrium
one can estimate for the unknown information transfer constant ratio
an expected range of at least $0.5\le\kappa_{r}/\kappa_{t}\le2$.
As long as one has no estimations from observations one can assume
$\kappa_{r}/\kappa_{t}\approx1$. However, the correct value of the
information transfer constant ratio, $\kappa_{r}/\kappa_{t}$, must
be determined from astronomical observations. For that the distance
and the age of an observed star must be well known. The radial velocity
must be negligible or also well known from galactic dynamics. Furthermore,
all known redshift corrections related to the properties/processes
of the star (like gravitational and mass-loss corrections \cite{Arp1992})
must be done and interactions of the propagating photons with cosmic
gas or cosmic dust in the line of sight must be negligible.

\section{Physical interpretation of the natural amount of equilibrium information\label{sec:Phys-interpretation}}

In appendix \ref{sec:Appendix-Signals} we explained from a general
point of view why the defined amount of information (equation (\ref{eqv3_01})),
which is related to a generic process variable, is a \emph{natural}
amount of information. We will now demonstrate by two physical examples
that the natural amount of equilibrium information, $I_{{\rm {eq}}}$,
has physical meaning. However, this approach requires detailed knowledge
or correct assumptions, respectively, of the internal/physical mechanisms
of the observed process.

\subsection{Moving ideal gas particles in a box}

A real gas reaches information equilibrium and becomes an ideal gas
if the gas volume becomes sufficiently large (see equation (\ref{eqv3_15}))
so that one has in information equilibrium 
\begin{equation}
I_{{\rm eq}}=2\frac{E}{\left|\delta E\right|}=f\frac{V}{\left|\delta V\right|}\label{eqv4_62}
\end{equation}
In thermodynamics ideal gas particles are approximated by point particles
(particles which have no own physical volume) so that the volume signal,
$\left|{\delta V}\right|$, is simply given by 
\begin{equation}
\left|\delta V\right|=\frac{V}{N}\label{eqv4_63}
\end{equation}
where $N$ is the number of ideal gas particles in a box of volume
$V$. That is, $\left|\delta V\right|$ is the average free volume
per gas particle. Considering equation (\ref{eqv4_62}) one gets 
\begin{equation}
I_{{\rm eq}}=f\cdot N\gg1\label{eqv4_64}
\end{equation}
so that the natural amount of equilibrium information is \emph{physically
realized} by the (huge) number $N$ of individual gas particles times
the constant factor $f$. We can also consider the following equation
\begin{equation}
i_{{\rm eq}}=\frac{I_{{\rm eq}}}{N}=f\label{eqv4_64a}
\end{equation}
and see that the amount of equilibrium information per gas particle
is given by the degree of freedom per gas particle, $f$.

Combining equations (\ref{eqv4_62}) and (\ref{eqv4_64}) one gets
for the energy signal 
\begin{equation}
\left|\delta E\right|=\frac{2}{f}\frac{E}{N}\label{eqv4_65}
\end{equation}
where $E$ is the internal energy of the moving ideal gas particles.
That is, $\left|\delta E\right|$ is proportional to the average energy
per gas particle. From thermodynamics it is well known that the internal
energy, $E\equiv U$, of an ideal gas is given by 
\begin{equation}
U=\frac{f}{2}Nk_{B}\, T\label{eqv4_66}
\end{equation}
where $k_{B}$ is the Boltzmann constant and $T$ the temperature
of the ideal gas. Combining equations (\ref{eqv4_65}) and (\ref{eqv4_66})
\begin{equation}
\left|\delta E\right|=k_{B}\, T\label{eqv4_67}
\end{equation}
one can conclude that the energy signal is proportional to the temperature.

How can we rationalize the process of moving ideal gas particles in
a box in the context of information transfer? Ideal gas particles
can be approximated by point particles so that energy is a generic
variable which is directly related to the ideal gas particles. The
(internal) energy is a \emph{restricted} process variable if the temperature
remains constant, so that one can identify the natural amount of information,
which is related to the energy, as the source of information \cite{Fielitz2011}.
Considering equations (\ref{eqv4_62}), (\ref{eqv4_64}) and (\ref{eqv4_67})
one gets for the amount of source information 
\begin{equation}
I_{E}^{{\rm source}}=2\frac{E}{\left|\delta E\right|}=2\frac{E}{k_{B}T}=f\cdot N=I_{{\rm eq}}\label{eqv4_68}
\end{equation}
which must be equal to the amount of equilibrium information. That
is, the gas particles and their (internal) energy is \emph{the source
of information} in the considered process. Furthermore, the ideal
gas particles are moving in a box of volume, $V$, so that there is
\emph{a second independent} generic variable $V$ related to the observed
process. According to equation (\ref{eqv3_01}) one can assign a natural
amount of information to this second independent process variable.
However, this natural amount of destination information, $I_{V}^{{\rm dest}}$,
related to the variable $V$, is at best equal to the natural amount
of source information, $I_{E}^{{\rm source}}$, which implies finally
equation (\ref{eqv4_62}), and hence, the ideal gas law.

\subsection{Diffusing tracer atoms in a homogeneous solid medium}

\begin{figure}
\includegraphics[width=8cm]{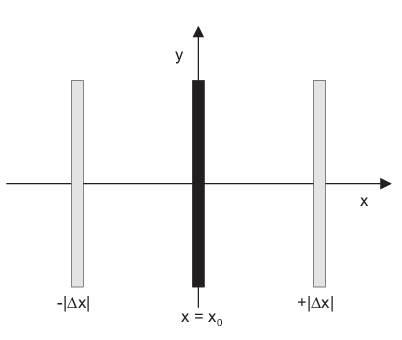} \protect\caption{In the plane $x=x_{0}$ of an isotropic homogeneous solid medium a
small layer of tracer particles was deposited at $t=t_{0}$. After
a given diffusion time interval, $\left|\Delta t\right|_{{\rm diff}}=t_{1}-t_{0}$,
one observes the same concentration of tracer particles located at
the distance $-\left|\Delta x\right|$ and $+\left|\Delta x\right|$
from the tracer particle source.}

\label{fig:04} 
\end{figure}

The microscopic details of the tracer diffusion process are well understood
so that we can now discuss the physical meaning of the natural amount
of equilibrium information. For that we start a thought experiment
and assume an isotropic homogeneous solid medium. In the first step
of our thought experiment we replace \emph{one} atom of the solid
medium by a tracer atom so that we can watch the movements of \emph{an
individual tracer atom}. We will observe that the individual tracer
atom makes short physical jumps with a given average jump frequency,
$\nu_{{\rm jump}}$. Because we assume an isotropic homogeneous solid
medium this observation is independent on the location in the solid
medium. It is, however, well known from diffusion theory (e.g. \cite{Mehrer2007})
that the jump frequency of the tracer atom depends on the temperature
of the diffusion medium, $\nu_{{\rm jump}}=\nu_{{\rm jump}}(T)$.
In the context of the information transfer concept the observed jump
frequency of the individual tracer atom corresponds to an observed
\emph{time signal} 
\begin{equation}
\left|\delta t\right|=1/\nu_{{\rm jump}}\label{eqv4_70}
\end{equation}
so that the total average number of physical jumps, $N_{{\rm jump}}^{{\rm obs}}$,
of an individual tracer atom during our observation time interval,
$\left|\Delta t\right|_{{\rm obs}}$, is given by 
\begin{equation}
N_{{\rm jump}}^{{\rm obs}}=\left|\Delta t\right|_{{\rm obs}}\nu_{{\rm jump}}\label{eqv4_71}
\end{equation}

In the next step of our thought experiment we deposit at low temperature
an amount of tracer particles in a small layer at $x=x_{0}$ inside
the isotropic homogeneous solid medium (see Fig.~\ref{fig:04}).
At low temperature the jump frequency of the deposited tracer particles
is negligible. At time $t=t_{0}$ we heat the solid medium quickly
up to a given temperature, $T$, so that all deposited tracer particles
of the diffusion source start to jump significantly with the same
average frequency, $\nu_{{\rm jump}}=\nu_{{\rm jump}}(T)$. At time
$t=t_{1}$ we observe again the tracer diffusion process and assume
that the observation time interval is much smaller than the diffusion
time interval, $\left|\Delta t\right|_{{\rm diff}}=t_{1}-t_{0}$.
In information equilibrium one has (writing equation (\ref{eqv4_07})
as an IE equation (\ref{eqv3_03})) 
\begin{equation}
I_{{\rm eq}}=2\frac{\left|\Delta t\right|_{{\rm diff}}}{\left|\delta t\right|}=\frac{\left|\Delta x\right|}{\left|\delta x\right|}\label{eqv4_69}
\end{equation}
Combining equations (\ref{eqv4_70}) -- (\ref{eqv4_69}) we get for
the natural amount of equilibrium information 
\begin{equation}
I_{{\rm eq}}=I_{t}^{{\rm source}}=2\left|\Delta t\right|_{{\rm diff}}\nu_{{\rm jump}}=2\cdot N_{{\rm jump}}^{{\rm diff}}\gg1\label{eqv4_72}
\end{equation}
The total average number of physical jumps, $N_{{\rm jump}}^{{\rm diff}}$,
during any diffusion time interval, $\left|\Delta t\right|_{{\rm diff}}$,
is equal for all tracer particles deposited at $t=t_{0}$ at $x=x_{0}$.
Because the time interval, $\left|\Delta t\right|_{{\rm diff}}$,
is \emph{restricted} one can identify the natural amount of information,
which is related to the time variable, principally as the source of
information \cite{Fielitz2011}. We now see that the natural amount
of equilibrium information is \emph{physically realized} by the (huge)
total average number of physical jumps, $N_{{\rm jump}}^{{\rm diff}}$,
of any individual tracer particle times the constant factor 2. We
can also consider the amount of equilibrium information per jump 
\begin{equation}
i_{{\rm eq}}=\frac{I_{{\rm eq}}}{N_{{\rm jump}}^{{\rm diff}}}=2\label{eqv4_72a}
\end{equation}
The factor 2 can be rationalized taking into account that after every
jump there are two \emph{options} with the \emph{same probability}:
the jump resulted in a (small) negative displacement in $x$ direction
or in a (small) positive displacement in $x$ direction, respectively.
This means that after any given diffusion time interval, $\left|\Delta t\right|_{{\rm diff}}$,
there is the same amount/concentration of tracer particles located
at the distance $-\left|\Delta x\right|$ and the distance $+\left|\Delta x\right|$
from the tracer particle source (see Fig.~\ref{fig:04}).

The distance, $\left|\Delta x\right|$, from the tracer particle source
is \emph{a second independent} generic process variable so that one
can assign a natural amount of information (equation (\ref{eqv3_01}))
to this independent variable. If the diffusion medium is sufficiently
large the length variable $x$ is an \emph{unrestricted} process variable
which is principally in the status of an information destination \cite{Fielitz2011}.
In the context of the information transfer concept one says that in
information equilibrium, $I_{x}^{{\rm dest}}=I_{t}^{{\rm source}}$,
during the diffusion time interval, $\left|\Delta t\right|_{{\rm diff}}$,
the natural amount of source information (quantified by equation (\ref{eqv4_72}))
has been completely transferred to the distance, $\left|\Delta x\right|$(which
is the information destination variable). In information equilibrium
one can evaluate the corresponding length signal, $\left|\delta x\right|$,
(combining equations (\ref{eqv4_69}) and (\ref{eqv4_72})) 
\begin{equation}
\left|\delta x\right|=\frac{1}{2}\frac{\left|\Delta x\right|}{N_{{\rm jump}}^{{\rm diff}}}\label{eqv4_73}
\end{equation}
which is proportional to the average displacement in $x$ direction
per jump.

\section{Summary}

In 2011 we presented a generalized concept of information transfer
\cite{Fielitz2011} which allows one to model information transfer
in natural systems. An important result of this concept is the global
IE theorem \ref{theo:01} where sufficient conditions are formulated
to reach information equilibrium for simple processes in nature. We
demonstrated the validity and the applicability of the global IE theorem
\ref{theo:01} on three well understood simple physical processes:
moving gas particles in a box, falling physical bodies on a planet
surface and propagating tracer particle from a point source. The (sufficient)
conditions to reach information equilibrium are quite general so that
one can expect that information equilibrium will occur quite often
in nature. One typical example of a process which is completely in
the state of information equilibrium (an EI process according to definition
\ref{def:03}) is the fall process of a small physical body on the
surface of a planet in vacuum. Another example is the tracer diffusion
process from an infinitesimally small tracer particle point source
in an infinite homogeneous medium. Such examples demonstrate that
the global IE theorem \ref{theo:01} formulates sufficient but not
necessary conditions for information equilibrium according to definition
\ref{def:01}.

The most interesting feature of the information transfer concept \cite{Fielitz2011}
is its \emph{system/process independent character}, so that it can
be helpful to get a new (phenomenological) point of view on a controversial
phenomenon. We showed in this paper that the information transfer
concept allows one to rationalize and to quantify the K-Trumpler effect
which is an unexplained controversial phenomenon since 1911 \cite{Arp1992}.
As discussed in section \ref{sub:Basics-Concluding} the information
transfer concept principally delivers quantitative but phenomenological
equations. That is, equation (\ref{eqv3_35}) is very useful for a
quantitative description of the K-Trumpler effect but can principally
not provide a deeper understanding of the internal/physical mechanisms
of light propagation from a star in vacuum on large scales. In this
context it should be recalled that the Hubble expansion is not expected
inside a galaxy, contrary to the K-Trumpler effect (confer \cite{Arp1992}
and references therein). That is, if the K-Trumpler effect is confirmed
more quantitatively by astronomical observations (equation (\ref{eqv3_35})
will be helpful for a quantitative tracking of this effect) it will
deliver new impulses to improve the theory of light propagation in
vacuum on \emph{large scales} (scales in the order of the diameter
of a galaxy and larger).

\appendix

\section{Signals and a natural amount of information\label{sec:Appendix-Signals}}

In this paper we consider an \emph{amount of information} and not
\emph{information per selection} as expressed by equation (\ref{eqv4_14})
or (\ref{eqv4_15}), respectively. Hartley \cite{Hartley1928} postulated
that the transmitted amount of information, $I$, is proportional
to the number, $n$, of selected symbols 
\begin{equation}
I=K\cdot n\label{eqv4_A1}
\end{equation}
where the postulated constant, $K=K(s)$, depends on the number $s$
of available symbols at each selection, and hence, on the operating
technical system which transfers information. According to Hartley's
own words ``the symbols convey by general agreement certain meanings
to the parties communicating.''

Based on Hartley's postulation we defined in \cite{Fielitz2011} an
amount of information, $I_{z}$, which is applicable to natural processes
\begin{equation}
I_{z}=\kappa_{z}\,\frac{\left|\Delta z\right|}{\left|\delta z\right|}\label{eqv4_A2}
\end{equation}
where $\left|\Delta z\right|$ is the \emph{absolute value} of the
generic process variable $z$ and $\kappa_{z}$ is an \emph{information
transfer constant}, related to the generic process variable $z$,
which depends on the considered natural system/process. Equation (\ref{eqv4_A2})
becomes descriptive if one assumes that $z$ is a length variable
so that we can model it by a thin rod of length $\left|\Delta z\right|$.
Next we can cut the thin rod into $n$ parts of equal length $\left|\delta z\right|$.
In this way we get $n$ equal rod parts (equal symbols). With our
communication partners we could arrange the meaning of the rod part
(the symbol), $\left|\delta z\right|$, for example, that one can
buy an apple. If we send any communication partner $n$ rod parts
every rod part (symbol), $\left|\delta z\right|$, would signal that
one can buy an apple so that our communication partner knows he/she
can buy $n$ apples. This analogue motivated us to call $\left|\delta z\right|$
the \emph{signal} of the process variable $z$. The discussed communication
example also demonstrates that our definition (\ref{eqv4_A2}) of
an amount of information implies principally a (very) simple communication
process where only one symbol/signal (a rod part of equal length $\left|\delta z\right|$)
is available per selection so that one has $s=1$. The simple case
$s=1$ is, however, also a special case because if one applies the
standard measure of information theory (equation (\ref{eqv4_14}))
one gets $I=n\cdot\ln(s)=n\cdot\ln(1)=0$, which tells us (wrongly)
hat an operating technical communication system which has only one
available symbol $(s=1)$ at each selection cannot transfer any information
at all. How can one solve this problem?

To answer the question let us first consider a \emph{traditional}
technical communication system with $s>1$. As an example we assume
that two symbols $(s=2)$ are available per selection: ``0'' and
``1''. Then we select for example 5 symbols and get for example
the string ``01011''. This is one example string of $N=s^{n}=2^{5}=32$
possible strings. The amount of transmitted information is in the
traditional sense $I=n\cdot\ln(s)=5\cdot\ln(2)$. Next we consider
a (very) \emph{simple} technical communication system where only one
symbol $(s=1)$ is available per selection, e.g. ``0''. Then we
select again for example 5 symbols and get the string ``00000''.
However, in this case it is the only possible string if we select
5 symbols $\left(N=s^{n}=1^{5}=1\right)$. The amount of transmitted
information is, therefore, in the traditional sense $I=n\cdot\ln(s)=5\cdot\ln(1)=0$,
which tells us (wrongly) that the considered simple technical communication
system could not transfer any information at all. Such a simple technical
communication system is, for example, realized if one has only black
balls available per selection $(s=1)$ and sends communication partners
boxes (the boxes are only for packing) with $n$ black balls. The
parties communicating by this simple technical communication system
would arrange the meaning of a black ball, for example, that a black
ball signals that one can buy an apple. If any communication partner
receives a box with $n$ black balls he/she knows that one can buy
$n$ apples. We see that there is information transfer possible by
such a simple technical communication system so that the standard/traditional
measures of information theory (equations (\ref{eqv4_14}) and (\ref{eqv4_15}))
have no meaning in the special/simple case $(s=1)$. To solve this
conflict one has to keep in mind that Hartley \cite{Hartley1928}
\emph{first postulated} that the transmitted amount of information
is proportional to the number $n$ of selected symbols (equation (\ref{eqv4_A1})).
(After that postulation he concluded that the postulated constant
should be $K=\ln s$.) That is, according to Hartley's postulation
the transmitted amount of information is \emph{primarily} proportional
to the number of selected symbols $n$. This means that in the considered
simple technical communication system (only black balls available
per selection) the transmitted amount of information is proportional
to the number $n$ of selected symbols (black balls).

Furthermore, one has to keep in mind that the logarithmic base of
the Hartley function, $K=\ln s$, is arbitrary so that the absolute
value of the postulated constant, $K$, in Hartley's postulation (\ref{eqv4_A1})
is a \emph{human-made} value. In contrast the postulated information
transfer constant, $\kappa_{z}$, in equation (\ref{eqv4_A2}) is
undefined in the context of the generalized information transfer concept
\cite{Fielitz2011}. That is, if one determines the information transfer
constant, $\kappa_{z}$, experimentally for a given natural process
its absolute value is fixed by the considered \emph{natural} process.
This motivated us to call the amount of information defined by equation
(\ref{eqv3_01}) explicitly the \emph{natural} amount of information.

\section{Comments related to the update of notations\label{sec:Appendix-Comments}}

\textbf{(a)} In \cite{Fielitz2011} we referred to the special case
$I_{x}=I_{y}$ as an ideal information transfer (IIT) process. This
was motivated by the derivation of theorems \ref{theo:01} and \ref{theo:02}
where an information transfer process from an information source to
an information destination was discussed. However, if we discuss only
the special case $I_{x}=I_{y}$, like in this paper, it is no longer
required to distinguish between an information source and an information
destination, so that it is more appropriate to refer to this special
case as an information equilibrium (IE).

\textbf{(b)} We referred to these theorems as IIT process theorems
in \cite{Fielitz2011}. This was motivated by the derivation procedure
where we considered ideal information transfer (IIT) processes. However,
finally these theorems formulate sufficient conditions to reach information
equilibrium (IE) so that it is more appropriate to refer to these
theorems as IE theorems.

\textbf{(c)} We referred to equation (\ref{eqv3_03}) as IIT process
equation in \cite{Fielitz2011}. With the new definition \ref{def:01}
it is more appropriate to refer to equation (\ref{eqv3_03}) as IE
equation.

\section{Derivation of equation (\ref{eqv3_29})\label{sec:Appendix-Derivation}}

The solution of the diffusion equation for an infinitesimally small
tracer particle point source in an infinite homogeneous medium is
given by (\cite{Crank1975} p. 29) 
\begin{equation}
C=\frac{M}{8(\pi D\,\left|\Delta t\right|_{{\rm diff}})^{3/2}}\,\exp\left(-\frac{r^{2}}{4\, D\,\left|\Delta t\right|_{{\rm diff}}}\right)
\end{equation}
where $M$ is the total amount of diffusing tracer particles and $D$
is the diffusion coefficient. The radial tracer particle flux, $j_{r}$,
is given by (e.g. \cite{Crank1975}) 
\begin{equation}
j_{r}=-D\frac{\partial C}{\partial r}
\end{equation}
With these equations one can calculate the local ratio $j_{r}/C$
and gets equation (\ref{eqv3_29}).


\begin{thebibliography}{10}
\bibitem{Fielitz2011}P. Fielitz and G. Borchardt, ``A generalized
concept of information transfer'', Physics Essays \textbf{24} (2011)
350.

\bibitem{Smith2013} J. R. Smith, $<$http://informationtransfereconomics.\\
blogspot.com/2013/04/an-informal-abstract.html$>$

\bibitem{Fielitz2014} P. Fielitz and G. Borchardt, ``The K-Trumpler
effect in the context of a generalized information transfer concept'',
Physics Essays \textbf{27} (2014) 365.

\bibitem{Fielitz2003} P. Fielitz, G. Borchardt, M. Schmücker and
H. Schneider, ``Silicon tracer diffusion in single crystalline 2/1-mullite
measured by SIMS depth profiling'', Phys. Chem. Chem. Phys. \textbf{5}
(2003) 2279.

\bibitem{Crank1975} J. Crank, \textit{The Mathematics of Diffusion}
(2nd ed., Oxford University Press, Oxford, 1975).

\bibitem{Arp1992}H. Arp, ``Redshifts of high-luminosity stars --
the K effect, the Trumpler effect and mass-loss corrections'', Mon.
Not. R. astr. Soc. \textbf{258} (1992) 800.

\bibitem{Hartley1928} R. V. L. Hartley, ``Transmission of Information'',
Bell System Tech. J. \textbf{7} (1928) 535.

\bibitem{Klir1998} G. J. Klir and M. J. Wiermann, \textit{Uncertainty-Based
Information: Elements of Generalized Information Theory} (Physica-Verlag,
Heidelberg, 1998).

\bibitem{Shannon1948} C. E. Shannon, ``A Mathematical Theory of
Communication'', Bell System Tech. J. \textbf{27} (1948) 379-423,
623-656.

\bibitem{Jaynes1957} E. T. Jaynes, ``Information Theory and Statistical
Mechanics'', Phys. Rev. \textbf{106} (1957) 620.

\bibitem{Haken2000} H. Haken, \textit{Information and Self-Organization:
A Macroscopic Approach to Complex Systems} (2nd ed., Springer, Berlin,
2000).

\bibitem{Frieden1998} B. R. Frieden, \textit{Physics from Fisher
Information: A Unification} (Cambridge University Press, Cambridge,
1998).

\bibitem{Mehrer2007} H. Mehrer, \textit{Diffusion in Solids - Fundamentals,
Methods, Materials, Diffusion-Controlled Processes} (Springer, Berlin,
2007).

\bibitem{Jones2007} B. W. Jones, \textit{Discovering the Solar System}
(2nd ed., John Wiley \& Sons Ltd., Chichester, 2007).

\bibitem{Carroll2007} B. W. Carroll and D. A. Ostlie, \textit{An
Introduction to Modern Astrophysics} (2nd ed., Addison-Wesley, San
Francisco, 2007).\end{thebibliography}
\end{document}